\documentclass[aps,pra,twocolumn,showpacs]{revtex4}
\usepackage{eurosym}
\usepackage{amsfonts}
\usepackage{amsmath}
\usepackage{amsopn}
\usepackage{amssymb}
\usepackage{graphicx}
\usepackage{graphics}
\usepackage{color}
\usepackage{verbatim}
\pdfoutput=1

\begin{document}

\title{Solitons in $\mathcal{PT}$-symmetric periodic systems with the quadratic
nonlinearity}
\author{F. C. Moreira$^{1,2}$, V. V Konotop$^{2}$, and B. A. Malomed$^{3}$}
\affiliation{$^{1}$Universidade Federal de Alagoas, Campus A. C. Sim\~oes - Av.
Lourival Melo Mota, s/n, Cidade Universit\'aria,
Macei\'o - AL 57072-900, Brazil}
\affiliation{$^{2}$Centro de F\'isica Te\'{o}rica e Computacional and Departamento de F\'isica, Faculdade de Ci\^encias, Universidade de Lisboa, Avenida
Professor Gama Pinto 2, Lisboa 1649-003, Portugal }
\affiliation{$^{3}$Department of Physical Electronics, School of Electrical Engineering,
Faculty of Engineering, Tel Aviv University, Tel Aviv 69978, Israel }
\date{\today}

\begin{abstract}
We introduce a one-dimensional system combining the $\mathcal{PT}$-symmetric
complex periodic potential and the $\chi ^{(2)}$
(second-harmonic-generating) nonlinearity. The imaginary part of the
potential, which represents spatially separated and mutually balanced gain
and loss, affects only the fundamental-frequency (FF) wave, while the real
potential acts on the second-harmonic (SH) component too. Soliton modes are
constructed, and their stability is investigated (by means of the
linearization and direct simulations) in semi-infinite and finite gaps in
the corresponding spectrum, starting from the bifurcation which generates
the solitons from edges of the gaps' edges. Families of solitons embedded
into the conttinuous spectrum of the SH component are found too, and it is
demonstrated that a part of the families of these \textit{embedded solitons}
(ESs) is stable. The analysis is focused on effects produced by the variation
of the strength of the imaginary part of the potential, which is a specific
characteristic of the $\mathcal{PT}$ system. The consideration is performed
chiefly for the most relevant case of matched real potentials acting on the
FF\ and SH components. The case of the real potential acting solely on the
FF component is briefly considered too.
\end{abstract}

\maketitle

\section{Introduction}

There is a growing interest in physical systems possessing the so-called $%
\mathcal{PT}$ (parity-time) symmetry~\cite{Bender_review,special-issues},
i.e., as a matter of fact, dissipative quantum systems with the antisymmetry
between spatially separated gain and loss. If the strength of the gain-loss
terms does not exceed a certain threshold value, the $\mathcal{PT}$
-symmetric system has a purely real spectrum and its non-Hermitian
Hamiltonians can be transformed into a Hermitian form \cite{Turkey}. Making
use of the similarity of the quantum-mechanical Schr\"{o}dinger equation to
the parabolic propagation equation in optics % \cite{Boyd2001},
it was proposed theoretically~\cite{Muga} and demonstrated experimentally
\cite{experiment} that the $\mathcal{PT}$ \ symmetry can be realized, in the
purely classical context of the wave propagation, in optics, where it
implies that a waveguide with the $\mathcal{PT}$-balanced gain and losses
allows the transmission of wave modes, emulating the index-guiding
transmission in ordinary (conservative) waveguides. These findings
stimulated numerous additional studies of the linear wave propagation in~\cite{special-issues} $%
\mathcal{PT}$-symmetric systems with particular attention being focused on the periodic potentials~\cite%
{PT_periodic} (see also review~\cite{review}).

Due to optical applications, additional interest has been recently attracted by nonlinear $\mathcal{PT}$ -symmetric optical systems with periodic modulation of the refractive index~\cite{Musslimani2008} which
demonstrated that stable solitons can be supported by the combination of the
Kerr nonlinearity and periodic complex potentials, whose spatially odd
imaginary part accounts for the balanced gain and loss. The stability of
such solitons was rigorously analyzed in Ref.~\cite{Yang}. Solitons can also
be naturally found in linearly-coupled dual-core systems with balanced gain
and loss in the two cores and intrinsic Kerr (cubic) nonlinearity in each
one \cite{dual}, and discrete solitons were predicted in coupled chains of $%
\mathcal{PT}$-symmetric elements \cite{discrete} and in general network of coupled $\mathcal{PT}$-symmetric oligomers (dimers, quadrimers, etc)~\cite{KPZ}.  
In addition to introducing
the usual Kerr nonlinearity, the $\mathcal{PT}$-symmetric part of the system
can be made nonlinear too, by introducing mutually balanced cubic gain and
loss terms ~\cite{AKKZ}. Properties of solitons in $\mathcal{PT}$ systems
may differ significantly from what is known about usual solitons in
conservative models. In particular, different families of solutions
bifurcating from different linear modes may merge in a single family,
exhibiting increased stability~\cite{Konotop2012}. On the other hand, the
increase of the gain-loss coefficient in the $\mathcal{PT}$-symmetric
Kerr-nonlinear coupler leads to shrinkage of the stability areas for $%
\mathcal{PT}$-symmetric and antisymmetric solitons, until they vanish when
this coefficient becomes equal to the inter-core coupling constant. We also mention recent intensive activity in study of the combined effect of linear and nonlinear $\mathcal{PT}$~\cite{combined} on existence and stability of optical solitons.

Apart from the Kerr nonlinearity, another fundamental type of nonlinear
interactions in optical media is quadratic \ ($\chi ^{(2)}$), which gives
rise to the second-harmonic-generation systems, which generate families of
two-color solitons,~\cite{chi2}. Recently the soliton dynamics in $\chi
^{(2)}$ materials was considered in the presence of a $\mathcal{PT}$%
-symmetric localized impurity~\cite{Moreira}. The objective of the present
work is to introduce a generic one-dimensional (1D) system with the $%
\mathcal{PT}$-symmetric periodic complex potential and conservative $\chi
^{(2)}$ nonlinearity, and construct stable solitons in it. The realization
of such a system in the spatial domain is quite possible in optics, using
appropriately juxtaposed gain and loss elements, like in Ref. \cite%
{experiment}, inserted into a $\chi ^{(2)}$ medium. We here focus on the
search for gap solitons (GSs) in the system with the periodic potential,
i.e., localized solutions whose propagation constant belongs to regions of
the forbidden propagation (\textit{gaps}) in the underlying linear spectra.
Similarly to the usual $\chi ^{(2)}$ systems (which do not include gain and
loss) \cite{chi2}, the quadratic nature of the nonlinearity makes the
interplay between the gaps of the fundamental-frequency (FF) and
second-harmonic (SH) fields a fundamental factor affecting GS families. In
particular, the generic mechanism of the creation of the families via
bifurcations from edges of the bandgaps \cite{Yang-book} can work in the FF
or SH component, or in both~\cite{Moreira}. We here analyze all these
possibilities.

It is relevant to mention that, as the $\mathcal{PT}$-symmetric systems is a
special type of settings at the border between conservative and dissipative
systems, the solitons that exist in them may be naturally compared not only
to their counterparts in conservative models (as mentioned above, concerning
the relation to GSs in the conservative $\chi ^{(2)}$ systems), but also to
solitons in generic dissipative systems, with unbalanced gain and loss. The
crucial difference of the dissipative solitons from their conservative
counterparts is that they exist, as isolated attractors of the system, at a
single value of the propagation constant, rather than continuous families
parametrized by an arbitrary propagation constant \cite{PhysD}. In
particular, as concerns GSs, 1D and 2D dissipative gap solitons in the
complex Ginzburg-Landau equations with periodic potentials were reported in
Ref.~\cite{CGL}. In this sense nonlinear $\mathcal{PT}$-symmetric systems, being not conservative and thus requiring the balance between dissipation and gain, but still allowing for existence of the continuous families of the solutions (what is the generic property of the such system provided the nonlinearity obey the same symmetry as the linear part~\cite{ZK}) occupy an intermediate position between the conservative and dissipative systems, and reduced to the former ones when the gain-loss coefficient becomes zero or the the later ones when appears dispalance between gain and loss.

The paper is organized as follows. The model is introduced in Sec. II. In
Sec. III basic results are reported for soliton families found in the model,
and the analysis of their stability, using both direct simulations and
linearized equations for small perturbations, is presented in Sec. IV. The
paper is concluded by Sec. V. Some special cases are separately considered
in two Appendices.

\section{The model}

We consider the $\chi ^{(2)}$ system, based on the evolution equations for
the FF and SH components, $u_{1}\left( \zeta ,\xi \right) $ and $u_{2}\left(
\zeta ,\xi \right) $, including the periodic $\mathcal{PT}$-symmetric
potential, with an imaginary component of amplitude $\alpha $, which is
assumed to act only onto the FF field:
\begin{subequations}
\label{final}
\begin{equation}
i\frac{\partial u_{1}}{\partial \zeta }=\frac{\partial ^{2}u_{1}}{\partial
\xi ^{2}}+\left[ V_{1}\cos (2\xi )+i\alpha \sin (2\xi )\right]
u_{1}+2u_{1}^{\ast }u_{2},  \label{final1}
\end{equation}%
\begin{equation}
i\frac{\partial u_{2}}{\partial \zeta }=\frac{1}{2}\frac{\partial ^{2}u_{2}}{%
\partial \xi ^{2}}+2\left[ V_{2}\cos (2\xi )+q\right] u_{2}+u_{1}^{2}.
\label{final2}
\end{equation}%
Here $\zeta $ and $\xi $ are the propagation and transverse coordinates, $q~$%
is the mismatch parameter, the $\chi ^{(2)}$ coefficient is scaled to be $1$%
, the asterisk stands for the complex conjugate, $V_{1}$ and $V_{2}$ are
amplitudes of the real part of the periodic potential for the FF and SH
components, while the period of the potential is set to be $\pi $ by means
of rescaling. Note that the conservative version of model (\ref{final}),
with $\alpha =0$, was studied in Refs.~\cite{Kartashov,Moreira1}, where stable solitons were found.

The most straightforward situation, which corresponds to the periodic
potential induced by a material grating etched into the $\chi ^{(2)}$
waveguide, corresponds to $V_{1}=V_{2}$ in Eqs.~(\ref{final}).
 Basic results are reported below for this situation. On the other
hand, a \textit{virtual grating }can be written in the waveguide by means of
the electromagnetically-induced-transparency mechanism, see, e.g., Ref. \cite%
{Guangzhou}. In the latter case, the effective periodic potential is
resonant, acting only in a narrow spectral interval. In this situation, it
is reasonable to consider the system with $V_{1}\neq 0$ and $V_{2}=0$, when
the potential does not affect the SH field, which is far detuned from the
resonance, and the opposite case, with $V_{2}\neq 0$ and $V_{1}=0$. To
illustrate similarities and differences between the different settings, some
results for the systems with $V_{2}=0$ (the virtual grating) and $%
V_{1}=V_{2}=0$ (the purely imaginary periodic potential) are presented in
Appendices A and B, respectively. In the latter case, the GSs are completely
unstable (as might be expected).

As concerns the loss and gain terms, they may be naturally assumed resonant
(e.g., if both are induced by resonant dopants, with the inverted and
uninverted populations in the gain and loss regions, respectively). For this
reason, it is natural to assume that these terms are present only in the FF
equation, as adopted in the system based on Eqs.~(\ref{final}). 
The opposite situation, with the imaginary potential acting on
the SH field, is possible too; it will be considered elsewhere.

We look for localized solutions with propagation constant $b$ in the form of
\end{subequations}
\begin{equation}
u_{l}\left( \xi ,\zeta \right) =w_{l}\left( \xi \right) e^{ilb\zeta },\quad
l=1,2,  \label{envelope}
\end{equation}%
where complex functions $w_{l}\left( \xi \right) $ obey the stationary
equations,
\begin{subequations}
\label{stat}
\begin{equation}
\frac{d^{2}w_{1}}{d\xi ^{2}}+\left[ V_{1}\cos (2\xi )+i\alpha \sin (2\xi )+b%
\right] w_{1}+2w_{1}^{\ast }w_{2}=0,  \label{stat1}
\end{equation}%
\begin{equation}
\frac{1}{2}\frac{d^{2}w_{2}}{d\xi ^{2}}+2\left[ V_{2}\cos (2\xi )+b+q\right]
w_{2}+w_{1}^{2}=0.  \label{stat2}
\end{equation}%

Generally speaking, Eqs.~(\ref{stat}) allow for solutions
obeying one of the following symmetries: $\{w_{1}(\xi ),w_{2}(\xi
)\}=\{w_{1}^{\ast }(-\xi ),w_{2}^{\ast }(-\xi )\}$ or $\{w_{1}(\xi
),w_{2}(\xi )\}=\{-w_{1}^{\ast }(-\xi ),w_{2}^{\ast }(-\xi )\}$.
Note also that, in the well-known cascading limit, $|q|\rightarrow\infty$
\cite{chi2}, Eq. (\ref{stat2}) yields $w_2\approx-w_{1}^{2}/\left( 2q\right)
$, and Eq. (\ref{stat1}) amounts to the equation with the cubic
nonlinearity,
\end{subequations}
\begin{equation}
\frac{d^{2}w_{1}}{d\xi^{2}}+\left[ V_{1}\cos(2\xi)+i\alpha\sin(2\xi )+b%
\right] w_{1}-q^{-1}\left\vert w_{1}\right\vert ^{2}w_{1}=0.  \label{casc}
\end{equation}
As mentioned above, solitons in the $\mathcal{PT}$ system based on Eq. (\ref%
{casc}) were recently studied in Refs.~\cite{Musslimani2008,Yang}.

\section{Gap-soliton families}

It is well known that $\chi ^{(2)}$ equations have particular solutions with
the vanishing FF component, $w_{1}\rightarrow 0$, while the SH part may
either vanish or remain finite. These solutions are usually subject to the
parametric instability\ \cite{chi2}, but they may be stabilized by an
additional cubic nonlinearity \cite{cubic}, by an external trapping
potential either~\cite{HS}, or by a $\mathcal{PT}$-symmetric localized
defect~\cite{Moreira}.

To identify bifurcations which give rise to GSs from edges of bandgaps, it
is also necessary to analyze the situation for $w_{1}\rightarrow 0$.
Generally speaking, one should then deal with three different cases~\cite%
{Moreira}, as shown below.

\noindent\emph{Case 1}: both components are of the same order,

\begin{equation}
w_{2}=\mathcal{O}(w_{1}),\quad w_{1}\rightarrow 0.  \label{case1}
\end{equation}

\noindent \emph{Case 2}: The SH field remains finite:
\begin{equation}
w_{2}=\mathcal{O}(1),~w_{1}\rightarrow 0.  \label{case2}
\end{equation}

\noindent \emph{Case 3}: The SH amplitude scales as the square of the FF
amplitude:
\begin{equation}
w_{2}=O(w_{1}^{2}),~w_{1}\rightarrow 0.  \label{case3}
\end{equation}%
Below, particular features of these three cases are considered separately.  

\subsection{Case 1}

\begin{figure}[!ht]
\begin{center}
\scalebox{0.5} {\includegraphics{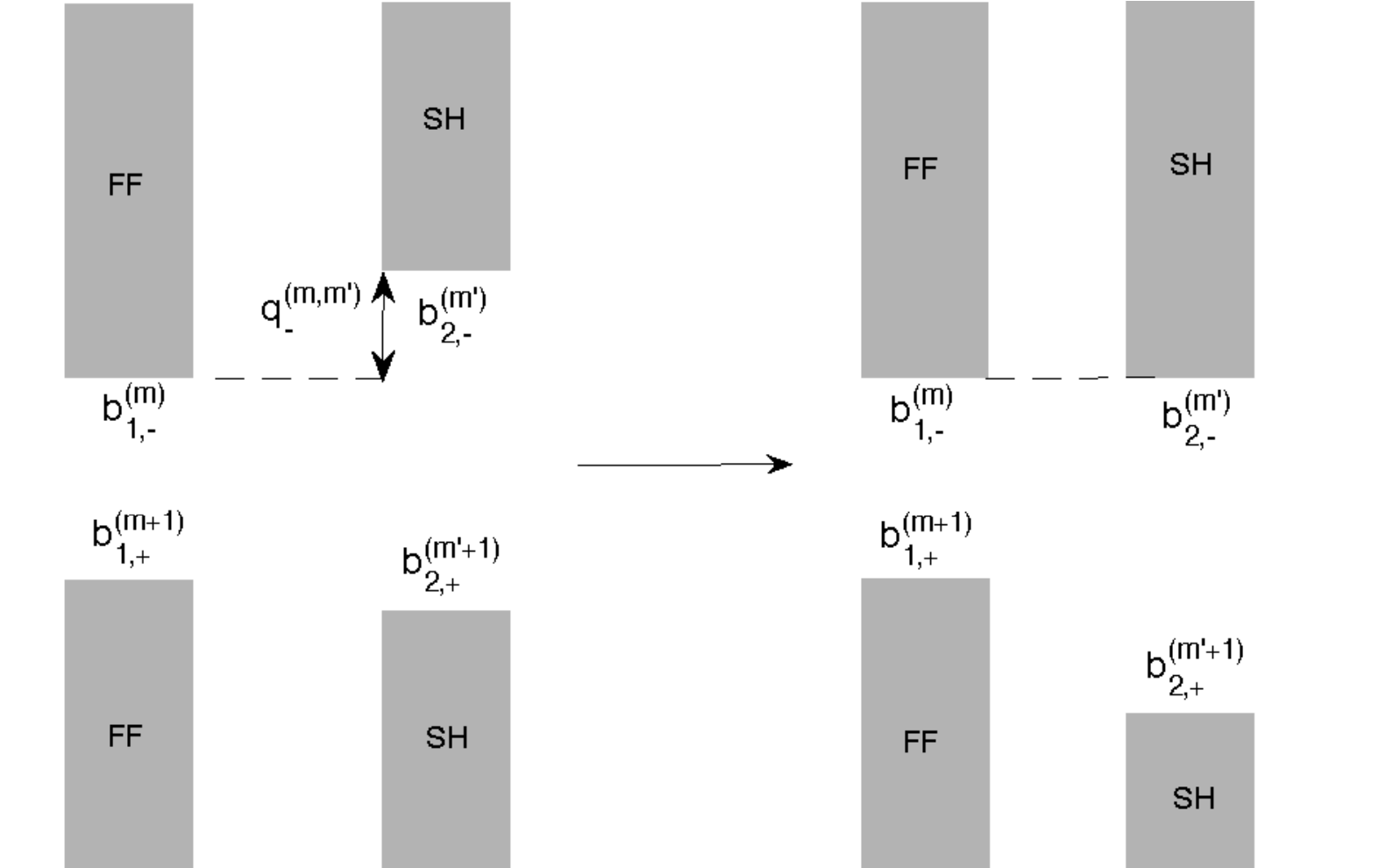}}
\end{center}
\caption{A schematic diagram illustrating matching the band edges of the FF and SH  for configuration of the Case 1. The left part of the figure represents the system without mismatch ($q=0$). The arrow in the middle shows to what configuration the band structure is transfered when the mismatch $q=q_{-}^{(m,m')}$, resulting in the existence of the total gap,  is imposed.}
\label{fig:qdef}
\end{figure}

In the limit case defined as per condition (\ref{case1}), the nonlinear
terms in both equations (\ref{stat}) can be neglected,
which, at the leading order, results in the system of decoupled linear
equations:
\begin{subequations}
\label{lin}
\begin{align}
\frac{d^{2}A_{1}}{d\xi ^{2}}+\left[ V_{1}\cos (2\xi )+i\alpha \sin (2\xi )+b%
\right] A_{1} =0,  \label{lin1} \\
\frac{d^{2}A_{2}}{d\xi ^{2}}+4\left[ V_{2}\cos (2\xi )+b+q\right] A_{2} =0.
\label{lin2}
\end{align}%
We notice that while Eq. (\ref{lin2}) is the well known Mathieu equation 
 the linear spectral problem (\ref{lin1}) with the $\mathcal{PT}$-symmetric
periodic potential was also thoroughly studied in literature~\cite{PT_periodic,periodic,Bender_review}. In particular, it is known that subject to constraint $|\alpha |\leq V_{1}$, 
equation (\ref{lin1}) gives rise to the pure real spectrum.
\begin{figure}[!ht]
\begin{center}
\scalebox{0.58} {\includegraphics{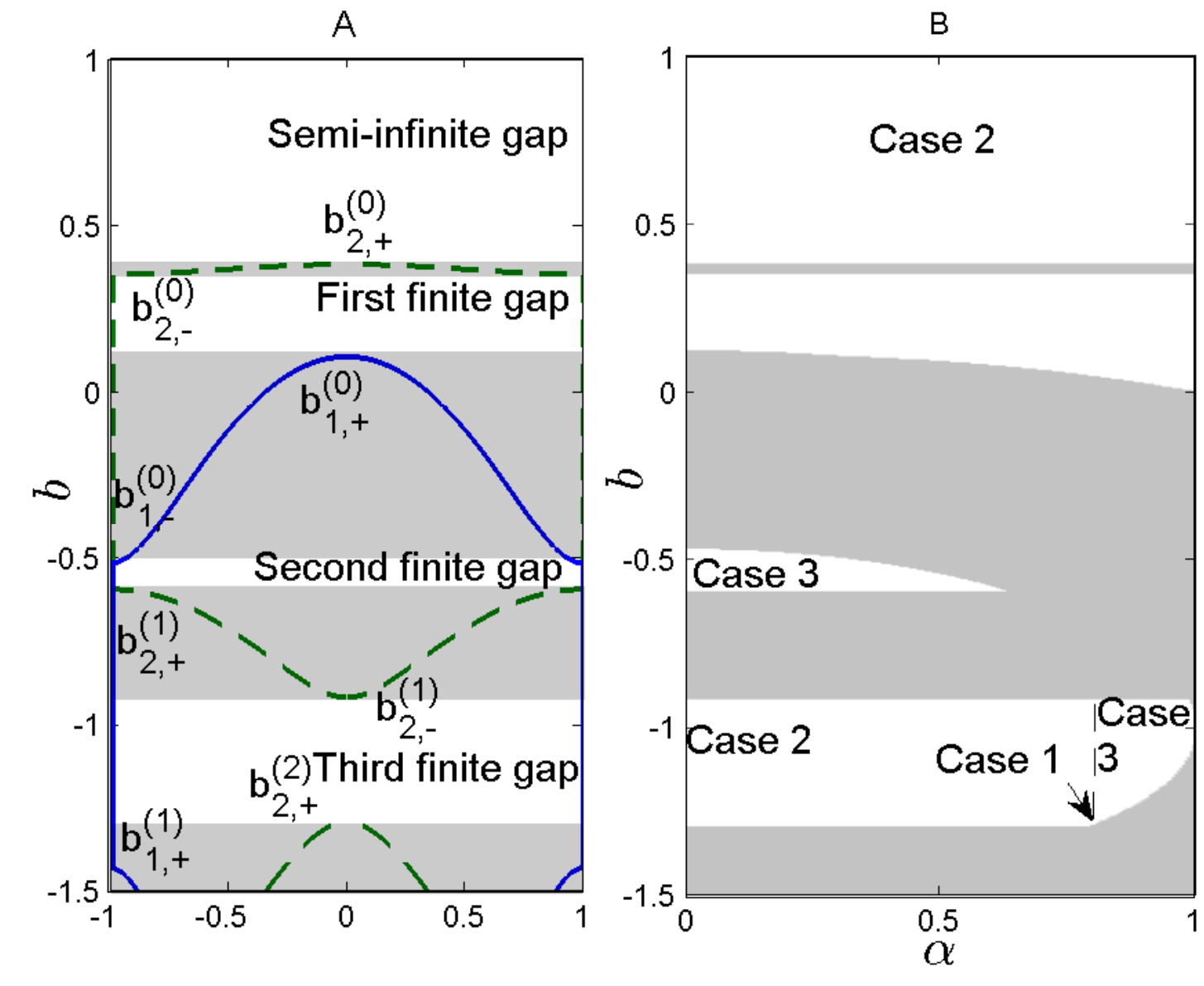}}
\end{center}
\caption{(Color online) Panel A: the spectrum of potential (\protect\ref{lin})
with $\protect\alpha =0.4$. The blue (solid) and green (dashed) curves
correspond to the FF and SH components, respectively. Regions of FF- and
SH-bands are shaded. The total gap correspond to white domains, as indicated
in the figure. Panel B: Propagation constant \emph{vs} the gain-loss
coefficient. Edges of the total gap determined by the SH component are
identifiable by horizontal lines, as they do not depend on $\protect\alpha $%
. The other edges are imposed by the FF component. The other parameters are $%
V_{1}=V_{2}=1$ and $q=0$. }
\label{fig:diagram}
\end{figure}
Now we turn to the \emph{combined} bandgap spectrum of
Eqs.~(\ref{lin}), i.e. to the values of the propagation constant $b$ which belong to the spectra of the both spectral problems. We denote by $b_{l,\pm }^{(m)}$ the propagation
constant at the upper ($+$) or lower ($-$) edge of the $m$-th
($m=0,1,2...$) band for the FF ($l=1$) and SH ($l=2$) components,
the latter being computed for $q=0$ (then the band edges of the SH
with $q\neq 0$ are given by $b_{2,\pm }^{(m)}-q$). Accordingly, the
sequence of finite gaps is defined as $\Sigma
_{l}^{(n)}=(b_{l,+}^{(n)},b_{l,-}^{(n-1)})$, where $n=1,2,...$, and
the semi-infinite gap is interval $\Sigma
_{l}^{(0)}=(b_{l,+}^{(0)},\infty )$. A \textit{total gap} is the
intersection of gaps of both components, as illustrated in panel A
of Fig.~\ref{fig:diagram}.

Condition (\ref{case1}) implies that in Eqs. (\ref{lin}), $b$ is a band edge
for the FF and SH components simultaneously (this situation is illustrated in the right panel of Fig.~\ref{fig:qdef}). Since the band edges of the FF 
and the SH are in general independent, to let case (\ref{case1}) occur, and hence
to let a branch bifurcate from the band edge $b_{1,\pm }^{(m)}$ of the FF, we
have to impose the following condition,
\end{subequations}
\begin{equation}
b_{1,\sigma }^{(m)}=b_{2,\sigma }^{(m^{\prime })}-q,\quad \sigma =\pm
\label{match}
\end{equation}%
where $m^{\prime }$ can be any band of the SH. Note, however, that only
edges of the same type allow the existence of the bifurcation we are dealing
with, which justifies the same sign, $+$ or $-$, on both sides of Eq. (\ref%
{match}). Indeed, as one can see in panel A of Fig.~\ref{fig:diagram}, the
individual gaps are located directly above (below) the band edges $%
b_{l,+}^{(m)}$ ($b_{l,-}^{(m)}$).
 
Once we impose condition (\ref{match}), we
force individual gaps to have at least one common edge. Then the total gap
only exists if this edge is either the lower or upper one for the both bands
simultaneously, as illustrated by Fig.~\ref{fig:qdef}. 

Condition (\ref{match}) imposes constraints on the design of the periodic
structure. Typically, $V_{l}$ would be fixed, and one could change the
concentration of the dopant, which amounts to varying the amplitude of the
imaginary part of the potential, $\alpha $, or mismatch $q$. Accordingly,
for given values values of $b_{1,\sigma }^{(m)}$ and $b_{2,\sigma
}^{(m^{\prime })}$, which are determined by the real part of the potential,
it is possible to satisfy Eq. (\ref{match}) by setting $q=q_{\sigma
}^{(m,m^{\prime })}$, where
\begin{equation}
q_{\sigma }^{(m,m^{\prime })}=b_{1,\sigma }^{(m)}-b_{2,\sigma }^{(m^{\prime
})},\quad \sigma =\pm .  \label{cond_q}
\end{equation}
All edges of the FF and the SH bands may be, in principle, matched with $%
q=q_{\sigma }^{(m,m^{\prime })}$. Additionally it is possible to match the
edges by tuning $\alpha $ alone, as can be seen in the third gap of panel B
in Fig.~\ref{fig:diagram} for $q=0$. It is also possible to see that Case 1
in the semi-infinite gap cannot be realized solely through adjusting $\alpha
$.
\subsubsection{Solitons in the semi-infinite gap}

\begin{figure}[!ht]
\begin{center}
\scalebox{0.5} {\includegraphics{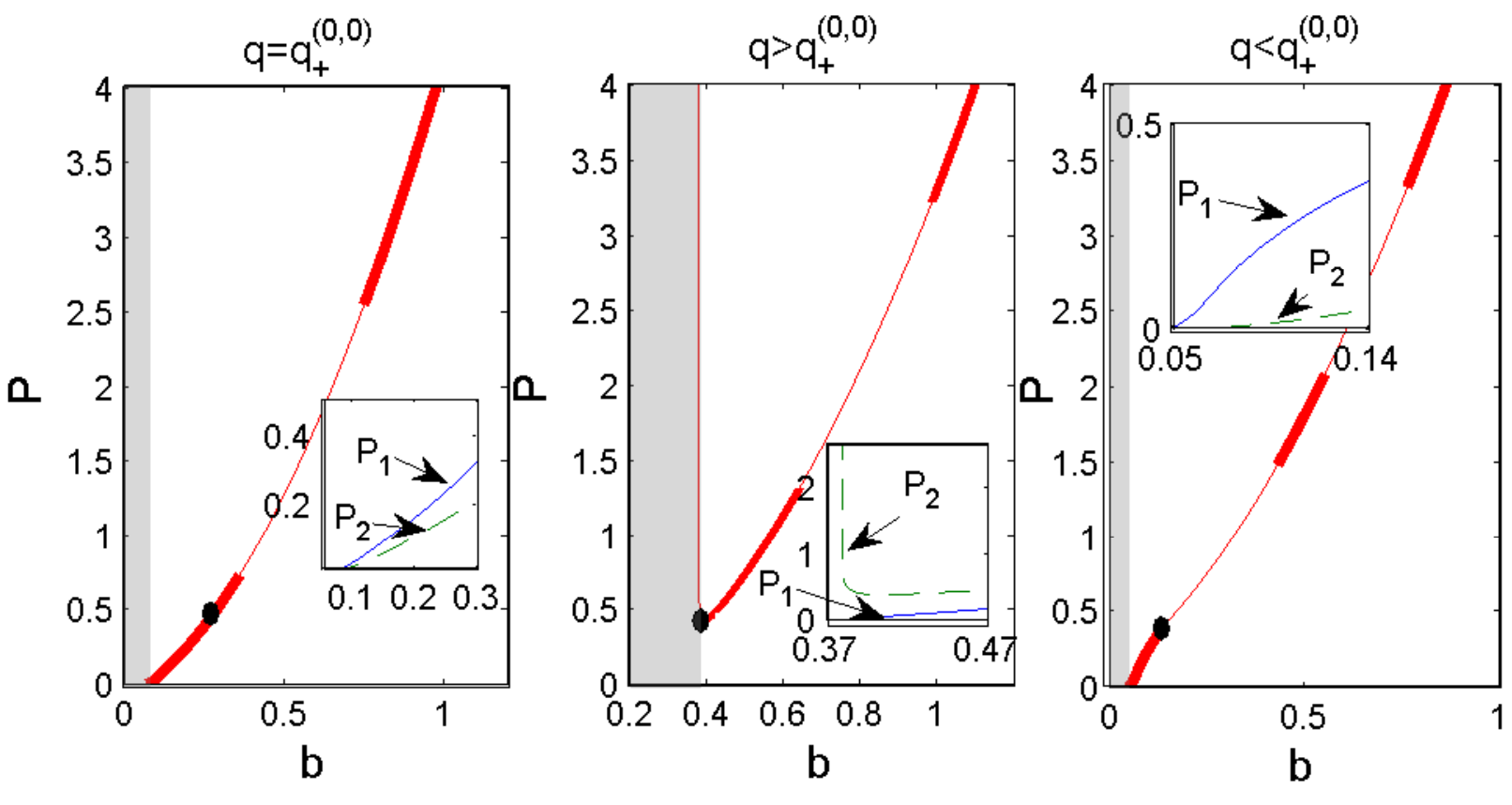}}
\end{center}
\caption{(Color online) Branches of fundamental solitons for $\protect\alpha %
=0.7$ and different values of $q$, found in the semi-infinite gap. The left,
central, and right panels correspond to cases 1, 2, and 3, with values $%
q=q_{+}^{(0,0)}=-0.316$, $q=0$ and $q=-0.5134$ respectively [see Eqs. (%
\protect\ref{case1}), (\protect\ref{case2}), and (\protect\ref{case3})].
Insets show power components $P_{1}$ (line) and $P_{2}$ (dashed line) close to an edge of the
semi-infinite total gap. Here and below, thick and thin lines represent
stable and unstable solutions, respectively. Shaded regions denote bands of
the FF and/or SH. Parameters are $V_{1}=V_{2}=1$ and $\protect\alpha =0.7$.}
\label{fig:branches}
\end{figure}

In Fig.~\ref{fig:branches}, we display branches of the
fundamental solitons, found numerically in the Case 1 in the semi-infinite gap, using matching $%
q=q_{+}^{(0,0)}$. The branches are presented in the plane ($b,P$), where  $P=P_{1}+P_{2}$, with $P_{l}\equiv l\int_{-\infty }^{\infty
}|w_{l}|^{2}d\xi $, ($l=1,2$), is the
total power.
\begin{figure}[!ht]
\begin{center}
\scalebox{0.58} {\includegraphics{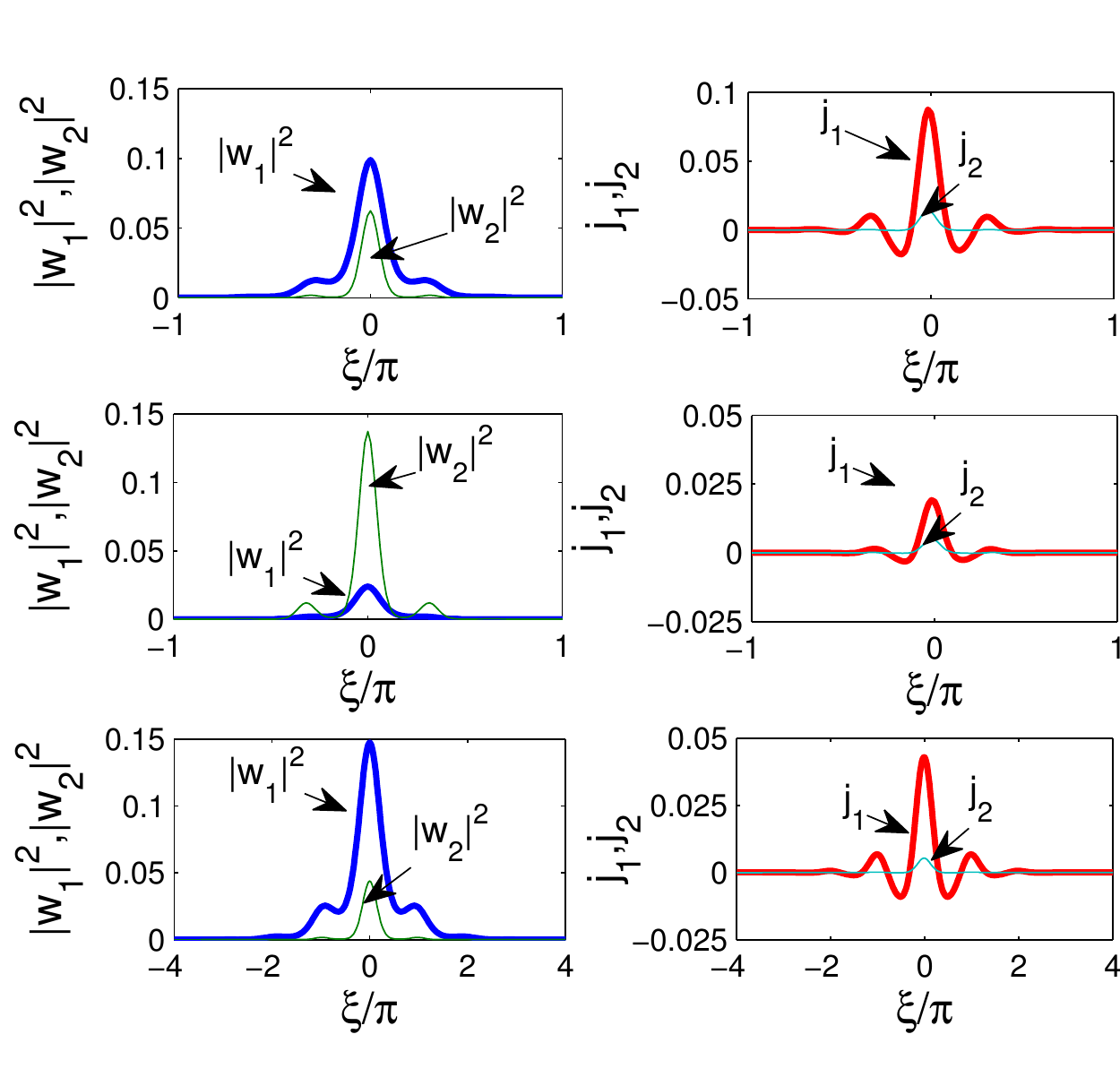}}
\end{center}
\caption{(Color online) Examples of stable fundamental solitons found in the
semi-infinite gap pertaining to all the three cases, which are indicated by
black circles in Fig.~\protect\ref{fig:branches}. The upper panels
correspond to case 1, with $b=0.25$ and band-edge matched with $%
q=q_{+}^{(0,0)}=-0.316$. The middle panels correspond to case 2, with $b=0.43$
and $q=0$. Lower panels show a solution of case 3 with $b=0.21$ and $q=-0.5134$%
. The parameters are $V_{1}=V_{2}=1$ and $\protect\alpha =0.7$.}
\label{fig:solutionssfP}
\end{figure}
Examples of fundamental soliton
solutions, i.e. the energy flows in each component as well as the currents are defined as
\begin{equation}
j_{l}(\xi)=|w_{l}|^2\frac{d \theta_{l}}{d\xi},
\qquad
\theta_{l}(\xi)=\text{arg}~w_l(\xi),
\end{equation}
corresponding to the total power $P=0.5$ are shown in Fig.~\ref{fig:solutionssfP}. 
We observe, that while the currents having maximum in the center and domains with alternating sign, have very similar shapes of the spatial profiles, the power density is mainly concentrated in the FF and SH in the Cases 3 and 2, respectively and is approximately equally split between the two components in the  Case 1 (as this is expected due to (\ref{case1})). % We also notice that in the Case 2, the amplitude of the current is an order of magnitude less than in the other two cases.
 In all three cases the real valued FF current $j_1$ has a significantly higher amplitude than the current of the SH, $j_2$, i.e. the balance between gain and losses is accomplished mainly due to the FF. 

Here and in the rest of the paper, localized solutions satisfying zero
boundary conditions were calculated numerically using a shooting method
described in detail in Ref.~\cite{Moreira1} for the conservative case, $%
\alpha=0$, and then extended to given $\alpha >0$ by means of the
Newton-Raphson method.

\subsubsection{Solitons in the third finite gap}

The main focus of this work is on the effects of the gain-loss coefficient $\alpha $ on branches of the fundamental solitons. To concentrate on this point, in what follows we set $q=0$.
\begin{figure}[!ht]
\begin{center}
\scalebox{0.55} {\includegraphics{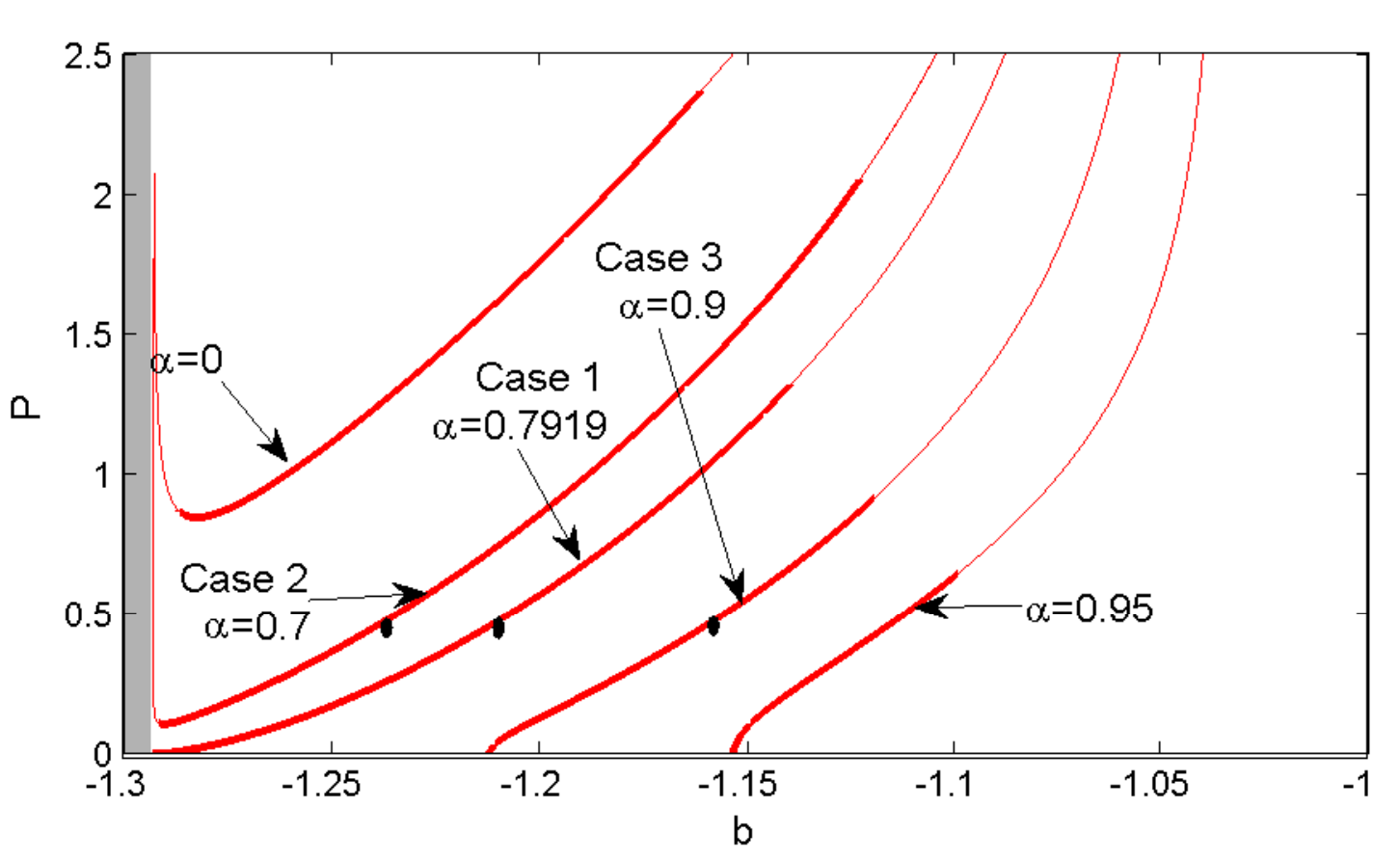}}
\end{center}
\caption{(Color online) Branches of fundamental GSs found in the third
finite gap for several values of amplitude $\protect\alpha $ of the
imaginary part of the periodic potential. All the three cases, 1, 2, and 3,
which are defined as per Eqs. (\protect\ref{case1}), (\protect\ref{case2}),
and (\protect\ref{case3}), respectively, are presented. The gray region
denotes the band of the SH component. The parameters are $V_{1}=V_{2}=1$, $%
q=0$.}
\label{fig:finbranch}
\end{figure}
For this choice it turns out possible to obtain the matching condition $%
b_{1,+}^{(1)}=b_{2,+}^{(2)}=1.29$ only in the third finite gap at $\alpha
=0.792$ (the value indicated by an arrow in Fig.~\ref{fig:diagram}B). In this context the  consideration of the third gap becomes particularly relevant, as one can examine
all three cases using only small deviations in parameter $\alpha$.
\begin{figure}[!ht]
\begin{center}
\scalebox{0.53} {\includegraphics{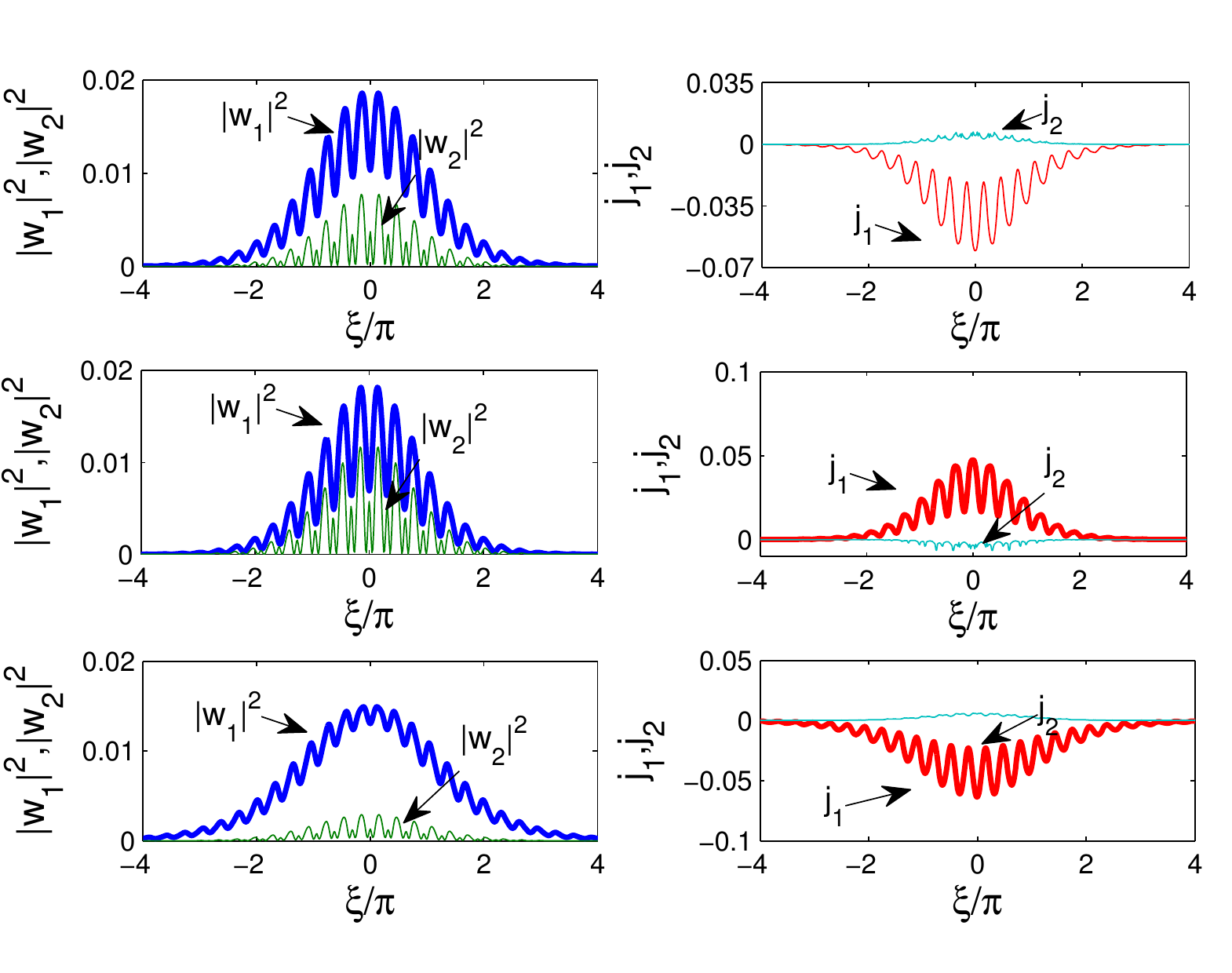}}
\end{center}
\caption{(Color online) Examples of stable fundamental gap solitons in the
third finite gap pertaining to all the three cases, which are indicated by
black circles in Fig. \protect\ref{fig:finbranch}. The upper panels
corresponds to Case 1, with $b=-1.207$ and band-edge matched with $\protect%
\alpha =0.7919$. The middle panels correspond to Case 2, with $b=-1.234$ and
$\protect\alpha =0.7$. Lower panels shows a solution of Case 3 with $b=1.553$
and $\protect\alpha =0.9$. The total power of all the three solitons is $%
P=0.5$. The parameters are $V_{1}=V_{2}=1$, $q=0$.}
\label{fig:solutionsP}
\end{figure}
The respective modifications of the branches subject to variation of the amplitude of the imaginary part of the potential $\alpha$ are illustrated Fig.~\ref{fig:finbranch}. Examples of  the profiles of the respective gap solitons are shown in Fig.~\ref{fig:solutionsP},  
where all three presented solutions have the same energy flow: $P=0.5$. The most significant distinction with the situation observed in Fig.~\ref{fig:solutionssfP} for the solitons in the semi-infinite gap is that (i) now the intensity of the FF is always bigger than the intensity of the SH and (ii) the energy currents of the FF and SH are counter propagating and having constant signs (the current $j_1$ is negative while $j_2$ positive). 
 
\subsection{Stability analysis}

The stability of the solutions found as outlined above was tested in direct
simulations, as well as within the framework of the linear stability
analysis.
 The later is based on the ansatz
\begin{equation}
u_{l}(\xi ,\zeta )=\left( w_{l}+\delta _{l+}e^{-i\lambda \zeta }+\delta_{l-}^{\ast }e^{i\lambda^{*}\zeta }\right) e^{ilb\zeta },
\label{pert}
\end{equation}
where $\delta _{l\pm }$ are amplitudes of small perturbations. The
substitution of this ansatz into Eqs.~(\ref{final}) leads to the linear
spectral problem,
\begin{equation}
\label{linp}
L\left(
\begin{array}{c}
\delta_{2+} \\
\delta_{1+} \\
\delta_{2-} \\
\delta_{1-}%
\end{array}%
\right) =\lambda \left(
\begin{array}{c}
\delta_{2+} \\
\delta_{1+} \\
\delta_{2-} \\
\delta_{1-}%
\end{array}%
\right)
\end{equation}%
where
\begin{equation}
L=\left(
\begin{array}{cccc}
L_{2}+2b & 2w_{1} & 0 & 0 \\
2w_{1}^{*} & L_{1}+b & 0 & 2w_{2} \\
0 & 0 & -L_{2}-2b & -2w_{1}^{*} \\
0 & -2w_{2}^{*} & -2w_{1} & -L_{1}^{*}-b%
\end{array}%
\right) ,
\end{equation}%
with 
\begin{subequations}
\begin{eqnarray}
L_{1} &=&\frac{d^{2}}{d\xi ^{2}}+V_{1}\cos (2\xi )+i\alpha\sin(2\xi), \\
L_{2} &=&\frac{1}{2}\frac{d^{2}}{d\xi ^{2}}+2\left[ V_{2}\cos (2\xi )+q%
\right] .
\end{eqnarray}%
Turning now to the stability properties of branches located in the semi-infinite  gap, we obtain that the fundamental branches may have one or more
instability intervals (see Fig.~\ref{fig:branches}). The lengths of these
intervals increase with $\alpha$ approaching the $\mathcal{PT}$-symmetry breaking point. This is a feature observed in all the cases considered below for semi-infinite gaps.
\begin{figure}[!ht]
\begin{center}
\scalebox{0.55} {\includegraphics{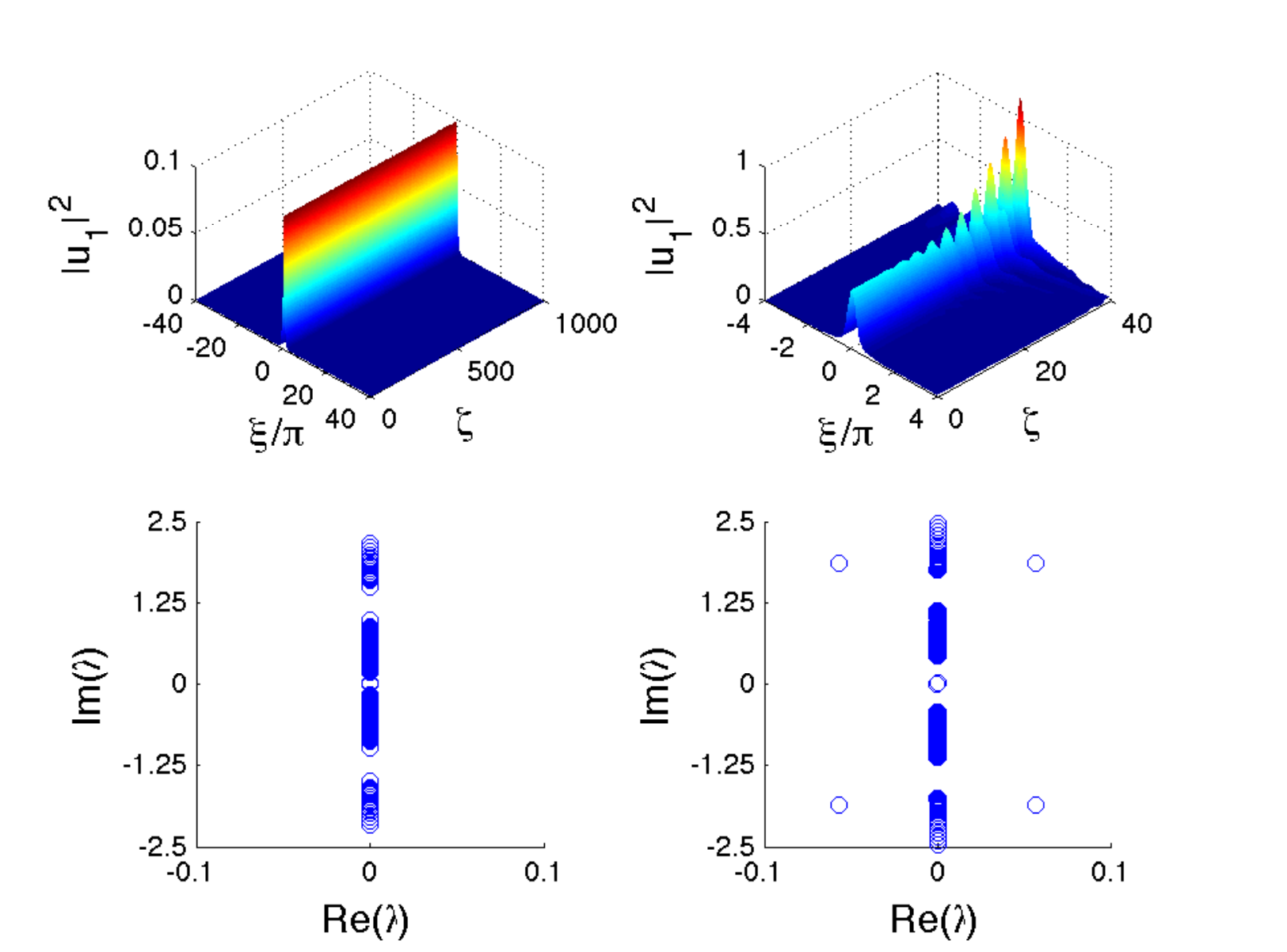}}
\end{center}
\caption{(Color online) Top plots: The evolution of two GS solutions with $%
10\%$ of amplitude random perturbations in Case 1 [see Eq. (\protect\ref%
{case1})] in the semi-infinite gap. Left panel has $b=0.25$ and is stable.
The right panel corresponds to unstable evolution of a solution with $%
b=0.5$. The corresponding eigenvalues of small perturbations are shown in
the lower panels. The parameters of the structure are $V_{1}=V_{2}=1$, $%
\protect\alpha =0.7$ and $q=q_+^{(0,0)}=-0.316$.}
\label{fig:propinfSIM}
\end{figure}
In Fig.~\ref{fig:propinfSIM} we show examples of the evolution of stable and unstable  localized solutions. The observed oscillatory instability is due to a quartet of complex $\lambda$ and instability develops as amplitude oscillations that increase with $\zeta$.
\begin{figure}[!ht]
\begin{center}
\scalebox{0.63} {\includegraphics{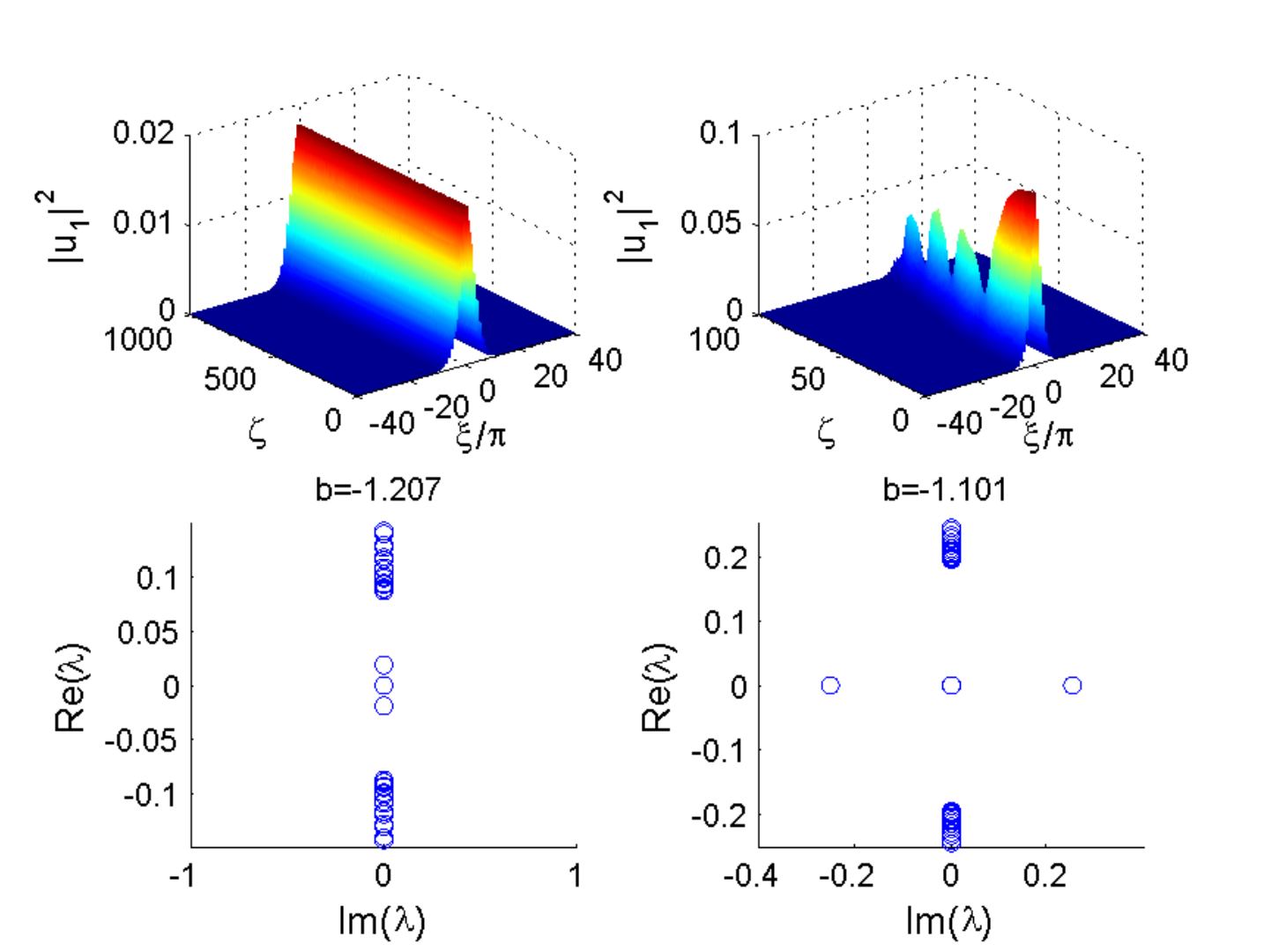}}
\end{center}
\caption{(Color online) Top plots: The evolution of two GS solutions with $%
10\%$ of amplitude random perturbations in Case 1 [see Eq. (\protect\ref%
{case1})] in the third finite gap. Left panel has $b=-1.207$ and is stable.
The right panel corresponds to unstable evolution of a solution with $%
b=-1.101$. The corresponding eigenvalues of small perturbations are shown in
the lower panels. The parameters of the structure are $V_{1}=V_{2}=1$, $%
\protect\alpha =0.7919$ and $q=q_+^{(0,0)}=0$.}
\label{fig:prop3finSIM}
\end{figure}

Stability of the solitons of the fundamental branches in the third finite gap (Case 1 with $\alpha=0.792$) is  shown in Fig.~\ref{fig:finbranch}. We observe an interval of stability which starts at the bifurcation point $b=b_{1,+}^{(1)}=b_{1,+}^{(2)}=1.29$, the rest of the branch corresponding to unstable solutions. 
 
Explicit examples of the direct propagation compared with the linear stability analysis are shown in Fig.~\ref{fig:prop3finSIM}. Stable and
unstable GSs with slightly modified $b$ belonging to the third finite gap are shown in the left and right columns, respectively. The two eigenvalues of the stable solution collide when $b$ is varied and assume purely imaginary values. It can be seen in the upper right panel that the perturbed solution decays very rapidly.

\subsection{Case 2}

Now we turn to numerical studies of solutions satisfying condition (\ref%
{case2}), in the vicinity of the total gap,
which coincides with an $m$-th SH\ gap edge, i.e. with $b_{2,\sigma }^{(m)}$. While FF component is vanishing in this case, i.e. $w_{1}\rightarrow 0$ as $b\rightarrow b_{2,\sigma }^{(m)}$, the amplitude of the SH $w_{2}$ persists finite while its width increases (i.e. the SH in this limit becomes delocalized).

In particular, the effect of the delocalization is responsible for the grows of the total power, i.e. divergence of $P$, at $b\to b_{2,+}^{(0)}$  and $\alpha =0.7$ shown in the central
panel of Fig.~\ref{fig:branches} and  at $b\to b_{2,+}^{(2)}$
shown  in Fig.~\ref{fig:finbranch} for the branches with $\alpha =0$ and $\alpha =0.7$ . 
Note that, in branches of Case 2 represented in Fig.~\ref{fig:finbranch} for
both $\alpha =0$ and $\alpha =0.7$, $P$ diverges at the same $%
b=b_{2,+}^{(2)} $, as the spectrum of Eq. (\ref{lin2}) is independent of $%
\alpha $. Similar results for the conservative system, with $\alpha =0$,
were previously obtained in~Refs.~\cite{Kartashov,Moreira1}. On the other
hand, no delocalization of the SH component was observed for the branches
satisfying condition (\ref{case2}) in~Ref.~\cite{Moreira}, where a $\mathcal{%
PT}$-symmetric localized potential was considered, since the bifurcation of the
second harmonic in that case departed from the localized defect state.
\begin{figure}[!ht]
\begin{center}
\scalebox{0.6} {\includegraphics{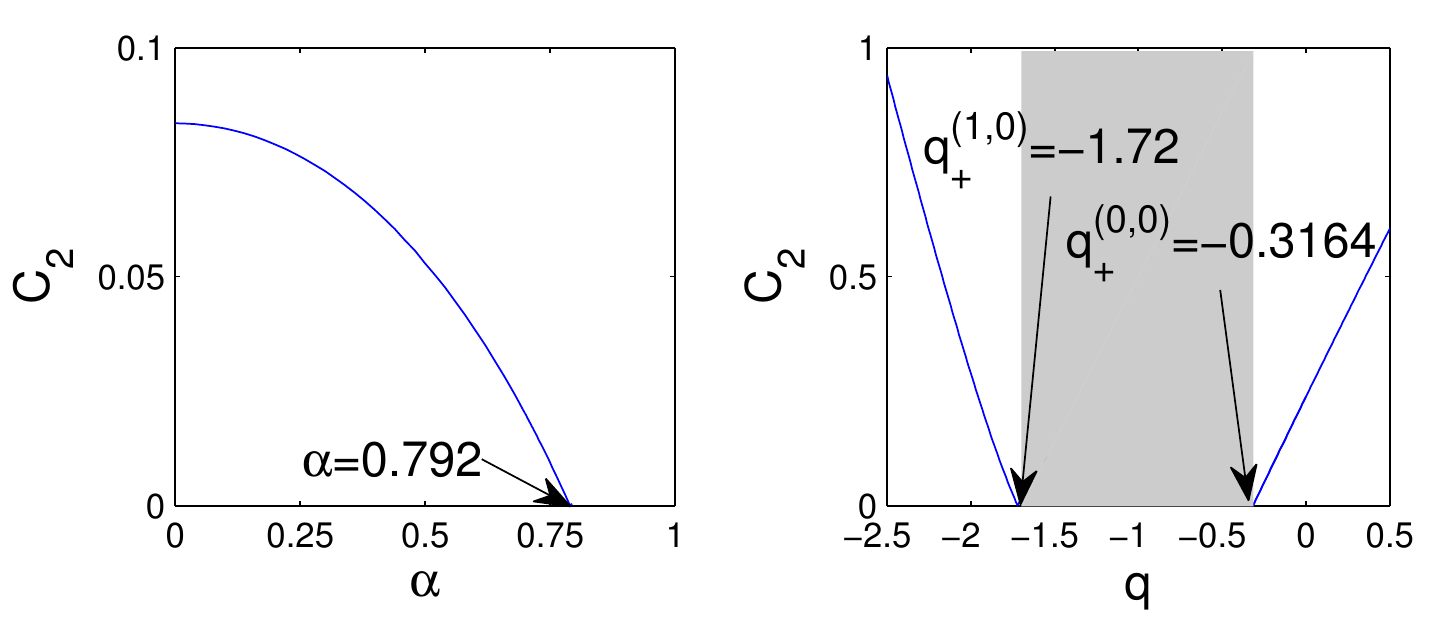}}
\end{center}
\caption{Left panel: $C_{2}$ \emph{vs.} $\protect\alpha $ at the SH edge $%
b=b_{2,+}^{(1)}=-1.293$ of the third finite gap for $q=0$. $C_{2}=0$ at $%
b_{1,+}^{(1)}=b_{2,+}^{(1)}$. Right panel:
 $C_{2}$ \emph{vs.} $q$ for $\protect\alpha =0.7$,   at the
edge $b=b_{2,+}^{(0)}=0.3784$. The shaded region represents the interval where $b=b_{2,+}^{(0)}+q$ falls inside the band $[b_{1,-}^{(0)},b_{1,+}^{(0)}]$. $C_{2}=0$ at $q=q_{+}^{(0,0)}$ and $q=q_{+}^{(1,0)}$. The parameters are $V_{1}=V_{2}=1
$.}
\label{fig:SHbif}
\end{figure}
The SH amplitude, we denote it by $C_{2}=\max {|w_{2}|%
}$, of a solution with $b$ close to $b_{2,\sigma }^{(m)}$, i.e., at $%
|b-b_{2,\sigma }^{(m)}|\ll 1$, depends on  the phase mismatch and on the gain-loss coefficient.

This is illustrated in the left panel of Fig.~\ref{fig:SHbif}, where we display plots $C_{2}$ {\it vs.} $\alpha$, calculated at the SH edge $b=b_{2,+}^{(1)}$ which coincides with the edge of the  third finite gap (like this is illustrated in the panel A of Fig.~\ref{fig:diagram}) for fixed $q=0$. In the right panel of Fig.~\ref{fig:SHbif} we show dependence of $C_2$ on the mismatch $q$ at the SH edge $b=b_{2,+}^{(0)}$ coinciding with the semi-infinite gap edge for fixed $\alpha=0.7$.  
We found that $C_{2}\rightarrow 0$ in the $C_2(\alpha)$ and $C_2(q)$ cases at specific value of the gain-loss coefficient: at $\alpha\approx 0.7919$ and $q=q_{+}^{(m,0)}$ respectively. In respect to values in which $C_2(q)\to 0$ we obtain this whenever $q$ just adjusts the edge of the SH band-edge, which in the present analysis is $b=b_{2,+}^{(0)}$, to be located exactly of an edge of the FF of the same type (in the figure this means it is a edge of upper type, $+$), exactly as described by formula (\ref{cond_q}). In Fig. \ref{fig:SHbif} we show only the values $q=q_{+}^{(0,0)}$ and $q=q_{+}^{(1,0)}$, which translates to the matching of edges $b_{1,+}^{(0)}=b_{2,+}^{(0)}+q_{+}^{(0,0)}$ and $b_{1,+}^{(1)}=b_{2,+}^{(0)}+q_{+}^{(1,0)}$.
In respect to the left panel of Fig.~\ref{fig:SHbif}, $C_2(\alpha)$ vanishes at the given value of the gain-loss coefficient corresponding  to the situation when the edges of the FF and SH gaps  
coalesce (i.e. $b_{1,+}^{(1)}=b_{2,+}^{(2)}$).

 Thus, whenever $C_2\to 0$ is attained by a proper choice of $\alpha$ or $q$, both the FF and SH components emerge with infinitely
small amplitudes $w_{1,2}$ when condition (\ref{match}) is met, i.e., when Case 2 transforms in Case 1. 
 
Examples of field profiles $w_{1,2}$ pertaining to the fundamental GS branches in the
semi-infinite and in the third finite bandgap are displayed in middle panels of Figs.~\ref{fig:solutionssfP} and \ref{fig:solutionsP}, respectively. In both figures the solutions are in the region close to the respective  SH band edges $b_{2,+}^{(0)}$ and $b_{2,+}^{(1)}$ where (\ref{case2}) is satisfied.
\begin{figure}[!ht]
\begin{center}
\scalebox{0.6} {\includegraphics{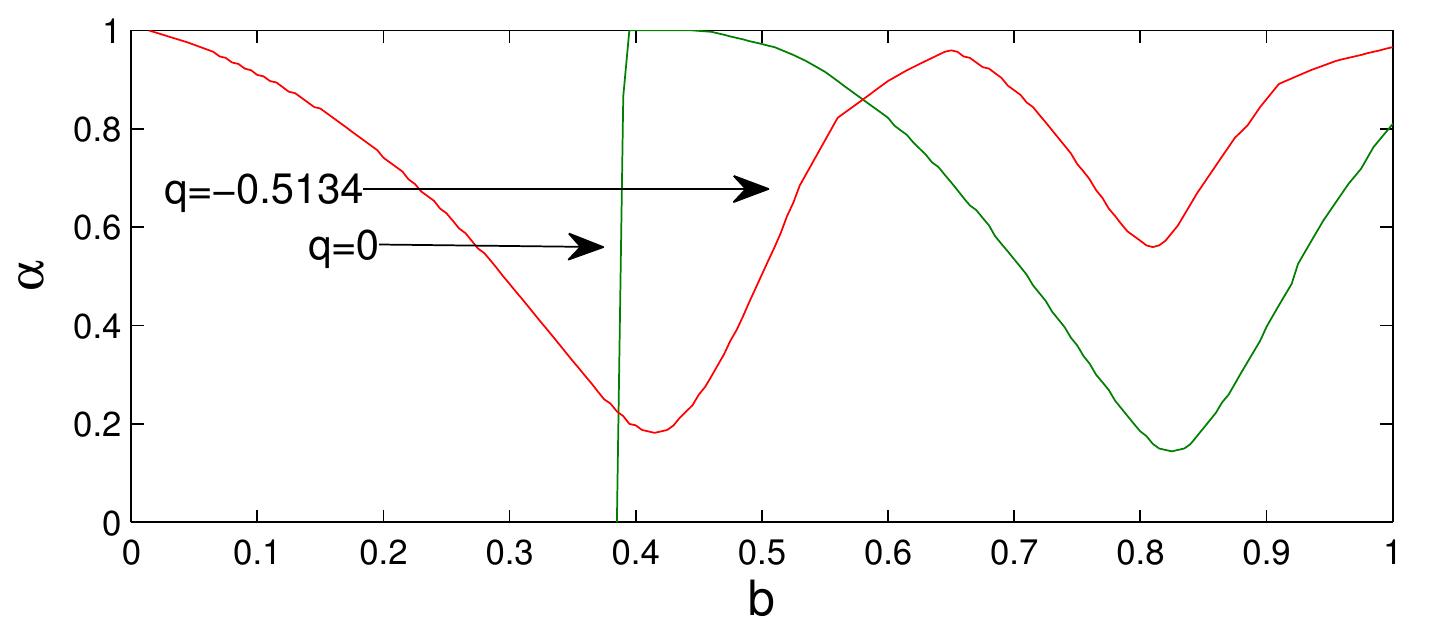}}
\end{center}
\caption{(Color online) The boundary between stable (below the curves) and unstable (above the curves) gap-soliton
solutions in the plane of $\left( b,\protect%
\alpha \right)$ obtained from the linear-stability analysis.   The curve with $q=0$ represents a Case 2 branch bifurcating from $b_{2,+}^{(0)}=0.3786$ and the curve with $q=-0.5134$ represents a Case 3 branch bifurcating from $b_{1,+}^{(0)}$. Note that while $b_{2,+}^{(0)}$ can be identified easily in the Case 2 curve as the point where the curve goes to $\alpha=0$, $b_{1,+}^{(0)}$ is not fixed because it depends on $\alpha$ (See panel B of Fig. \ref{fig:diagram}). The parameters are $V_{1}=V_{2}=1$.}
\label{fig:stability1}
\end{figure}

As concerns the stability of the GSs, Case 2 has one notable difference in
  the semi-infinite and in the third finite gaps in comparison with Case 1,
 whenever a given branch satisfying (\ref{case2}) bifurcates from a SH edge $b_{2,\sigma }^{(m)}$,  a small unstable region close to $b_{2,\sigma }^{(m)}$ that persists even when $\alpha =0$ exists.
\begin{figure}[!ht]
\begin{center}
\scalebox{0.6} {\includegraphics{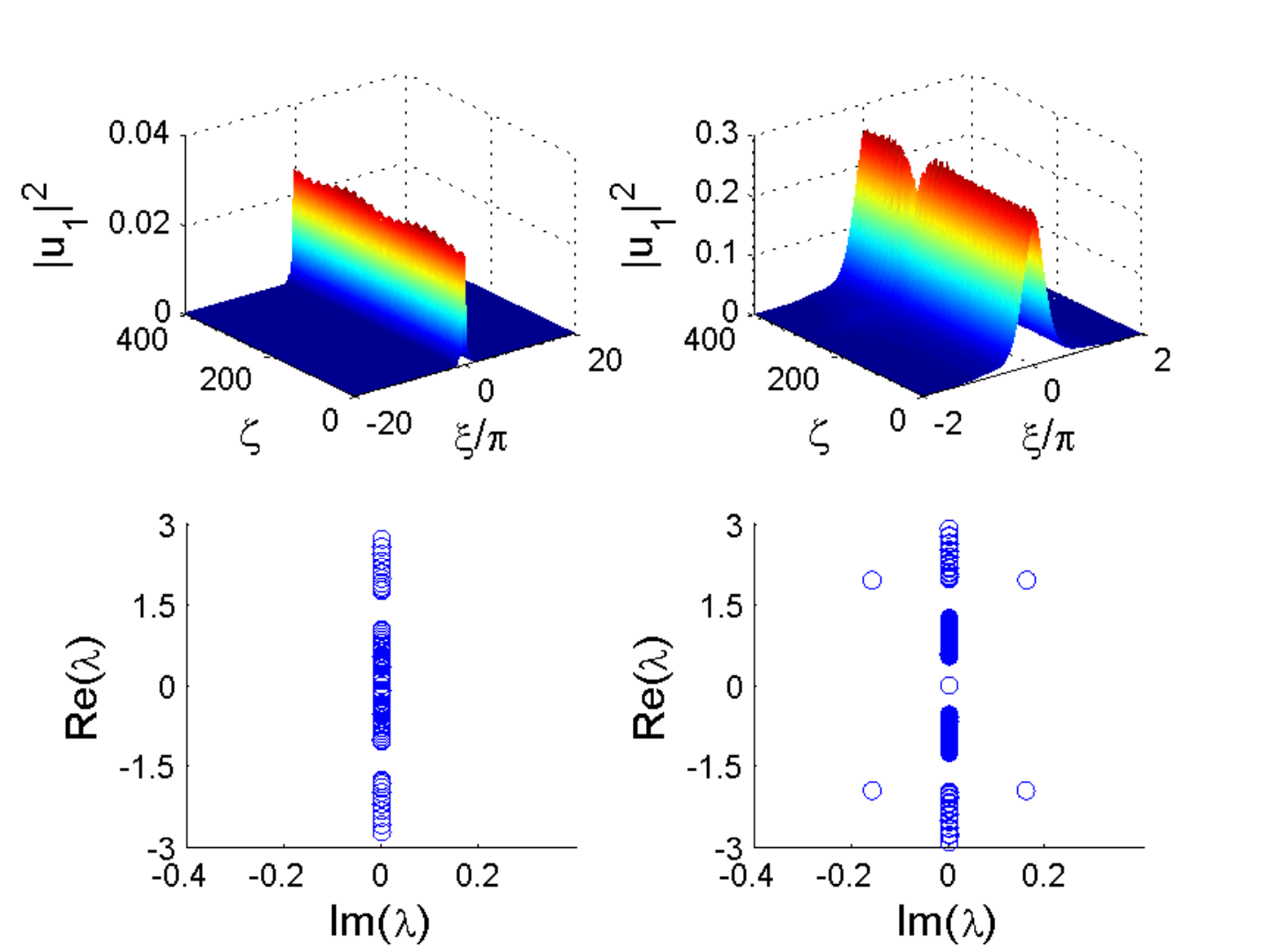}}
\end{center}
\caption{(Color online) Top plots: The evolution of two GS solutions with $%
20\%$ of amplitude random perturbations in Case 2 [see Eq. (\protect\ref%
{case2})] in the semi-infinite gap. Left panel has $b=0.4$ and is stable.
The right panel corresponds to unstable evolution of a solution with $%
b=0.6$. Note that the linearly unstable solution remains localized. The corresponding eigenvalues of small perturbations are shown in
the lower panels. The parameters of the structure are $V_{1}=V_{2}=1$, $%
\protect\alpha =0.7$ and $q=0$.}
\label{fig:frontierw-084}
\end{figure}
In Fig. \ref{fig:stability1} the curves separating stable and unstable solutions of the fundamental branch values of $q=0$ and $q=-0.5134$ are shown in the plane $(b,\alpha)$. The Case 2 branch is the curve with $q=0$ in the semi-infinite gap, where can be seen stability threshold abruptly decays to zero. The other curve with $q=-0.5134$ a Case 1 bifurcation, it do not share this property. In both curves is possible to see that, as we reported in the previous section, there may be one or more unstable intervals with lenghts that increase with $\alpha$.

Examples of the propagation of
stable and unstable solutions with variations in $b$ in the semi-infinite gap are shown in Fig. \ref{fig:frontierw-084}. Instability appear due to the collision of internal modes with the band edges of the spectrum of (\ref{linp}) resulting in four complex eigenvalues $\lambda$. The propagation however shows that the perturbed solution can remain localized despite amplitude oscillations as is possible to see in the upper left panel of Fig. \ref{fig:frontierw-084}. At about $\zeta=300$ there is an emission of energy from the localized field region but the structure quickly regains energy and remains localized.
\begin{figure}[!ht]
\begin{center}
\scalebox{0.6} {\includegraphics{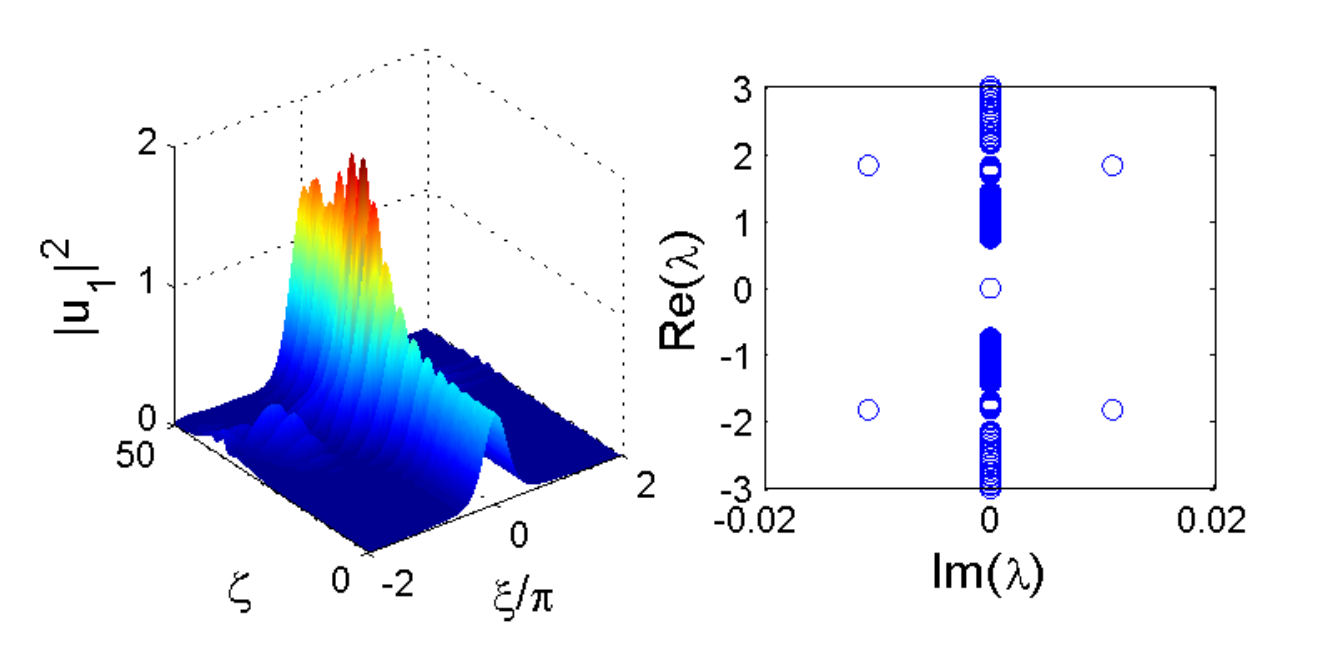}}
\end{center}
\caption{(Color online) Left plot shows the evolution of an unstable localized solution with $b=0.8$ added by $%
10\%$ of amplitude random perturbations in Case 2 [see Eq. (\protect\ref%
{case2})] in the semi-infinite gap. The right plot shows the corresponding
eigenvalues of small perturbations. The parameters of the structure are $V_{1}=V_{2}=1$, $\alpha=0.7$ and $q=0$}.
\label{fig:b08alpha07q0SH}
\end{figure}
However not all linearly unstable solutions have this behaviour. 
\begin{figure}[!ht]
\begin{center}
\scalebox{0.55} {\includegraphics{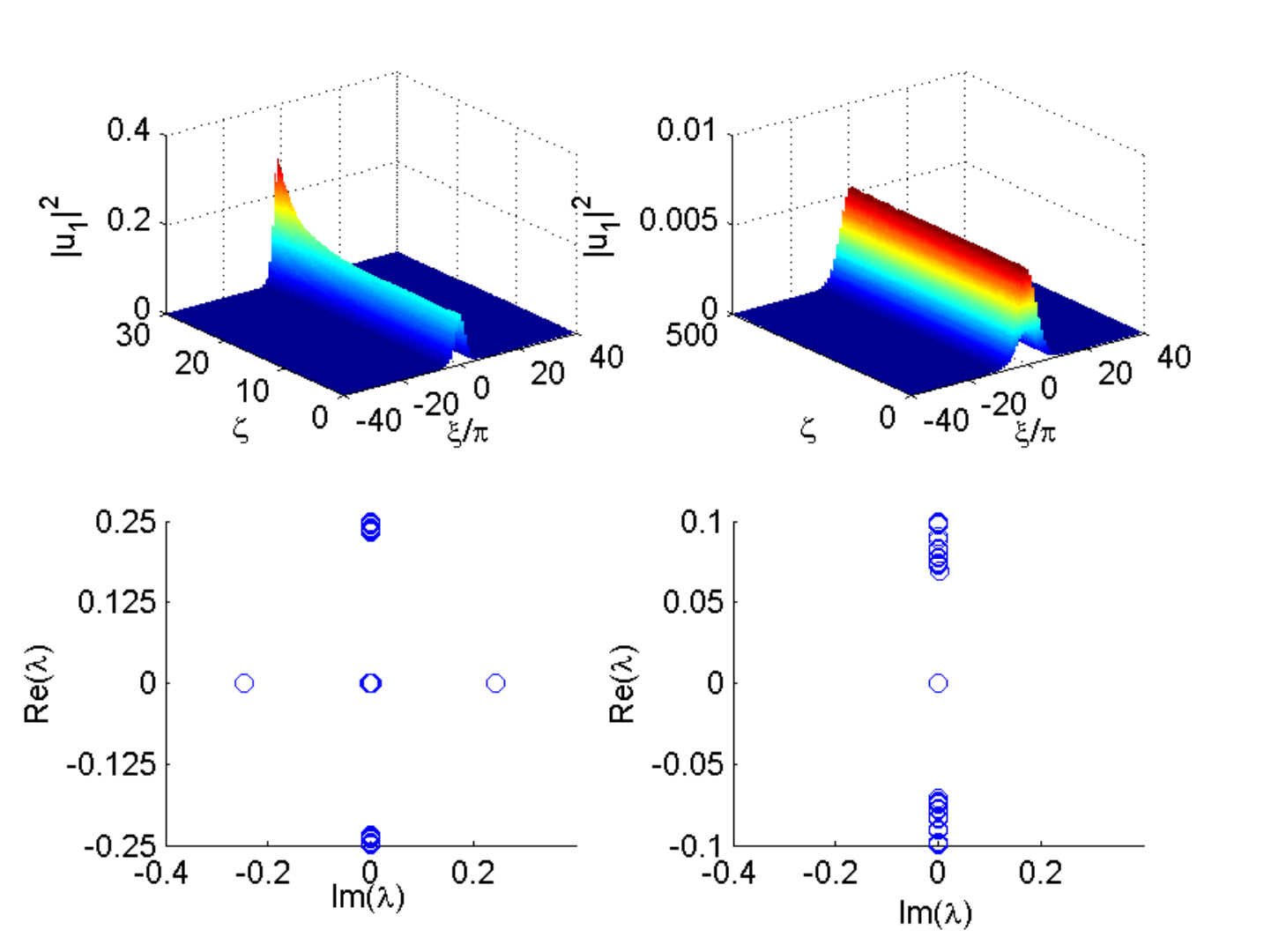}}
\end{center}
\caption{(Color online) Top plots: The evolution of two GS solutions with $%
10\%$ of amplitude random perturbations in Case 2 [see Eq. (\protect\ref%
{case2})] in the third finite gap. Left panel has $b=-1.11$ and is unstable.
The right panel corresponds to stable evolution of a solution with $%
b=-1.268$. The corresponding eigenvalues of small perturbations are shown in
the lower panels. The parameters of the structure are $V_{1}=V_{2}=1$, $%
\protect\alpha =0.7$ and $q=0$.}
\label{fig:prop3finSH}
\end{figure}
In Fig. \ref{fig:b08alpha07q0SH} we show an unstable solution that rapidply decays.

Examples of solutions in the third finite gap with slightly different $b$ are shown in Fig. \ref{fig:prop3finSH}. Unstable eigenvalue with positive $\lambda$ appear when $b$ is slightly bigger than $b=-1.268$ of the stable solution. Instability develops as a rapid increase of the amplitudes of the intensities $|w_{1,2}|^2$ with propagation.
\subsection{Case 3}

Finally, we consider GS branches generated by bifurcations which obey
condition (\ref{case3}) satisfied in a vicinity of the FF edge of the total
gap, $b=b_{1,\sigma }^{(m)}$.
\begin{figure}[!ht]
\begin{center}
\scalebox{0.55} {\includegraphics{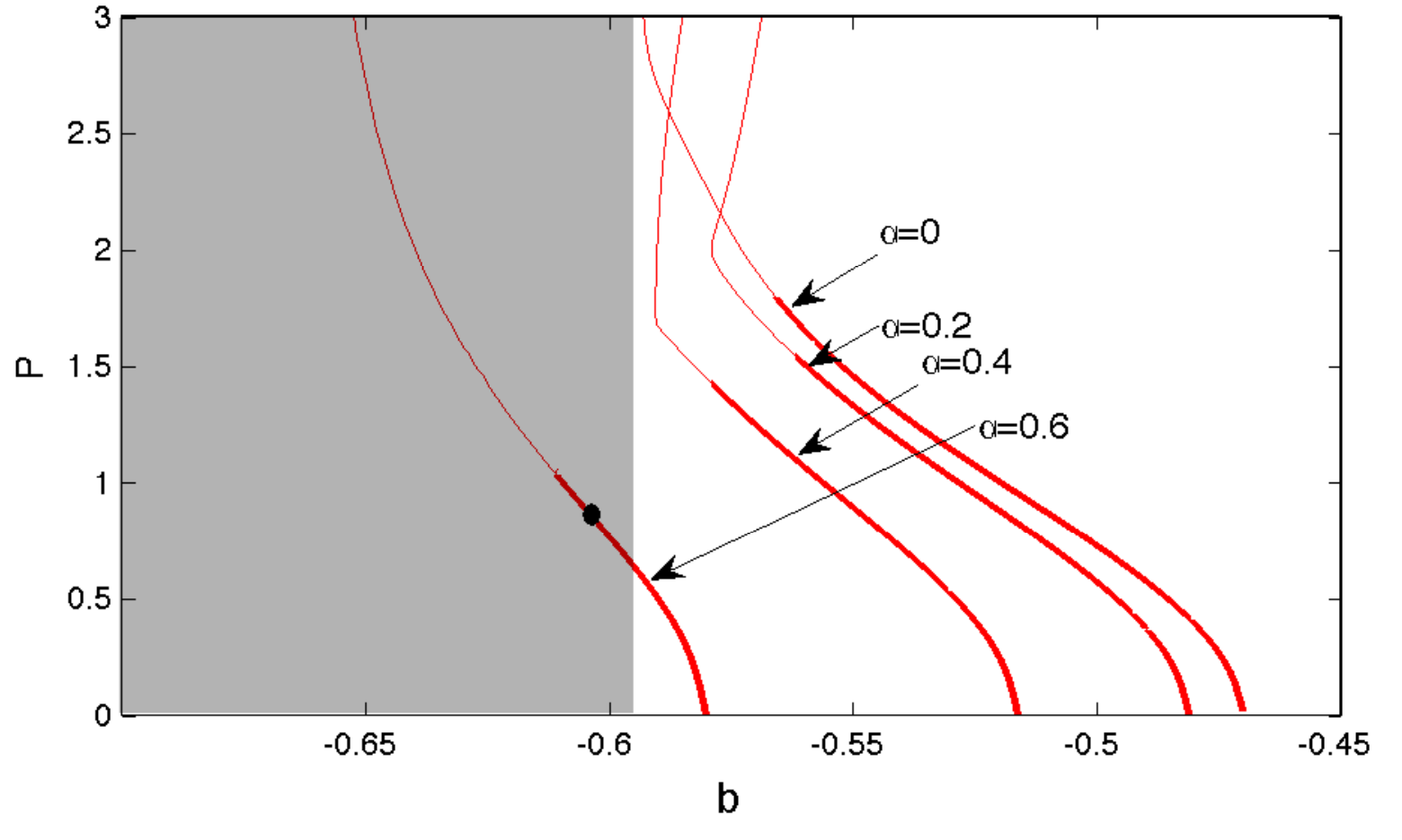}}
\end{center}
\caption{(Color online) Branches of fundamental GSs for several values of $%
\protect\alpha $ in the second finite gap. The bifurcations are of the Case
3 type. The shaded region denotes the band of the SH component. The branches
extend into the SH band, as \textit{embedded solitons}, atr $\protect\alpha %
>0.41$. Thick and thin lines represent stable and unstable solutions,
respectively. The parameters are $V_{1}=V_{2}=1$, $q=0$.}
\label{fig:branchband}
\end{figure}
The branch of the fundamental GS solutions pertaining to Case 3
is displayed in the right panel of Fig. \ref{fig:branches} for the
semi-infinite gap.

An example of GS solution is displayed in the lower panels Fig. \ref{fig:solutionssfP}. It can be seen that $|w_1|^2$ has a much higher amplitude than $|w_2|^2$. Fundamental branches of Case 3 are also represented by the branches with $\alpha =0.9$
and $\alpha =0.95$ in Fig. \ref{fig:finbranch} for the third gap. An example
of the respective field profiles for the GS branch in the third finite
bandgap, bifurcating from $b_{1,+}^{(1)}$, is shown in the lower panels of Fig. \ref{fig:solutionsP}. Also in the same figure, can be noted that while $|w_1|^2$ is strictly positive, $j_1$ is strictly negative. The current $j_2$ is strictly positive.

We observe in Fig. \ref{fig:branchband}, where branches for several values
of $\alpha $ are found in the second finite gap, that the branch which
bifurcates from $b_{1,+}^{(0)}$ at $\alpha =0.6$ goes into the band of the
SH, where it becomes a family of \textit{embedded solitons} (ESs) \cite%
{Embedded,embed}, i.e., those existing inside (\textit{embedded into}) the
continuous spectrum. The existence of such solitons is explained by fact
that their decaying asymptotic tails at $|\xi |\rightarrow \infty $ follow
relation (\ref{case3}), hence the SH equation is \textit{non-linearizable}
for the decaying tails, invalidating the standard argument for the
non-existence of solitons whose propagation constant falls into the band.
\begin{figure}[tbp]
\begin{center}
\scalebox{0.5} {\includegraphics{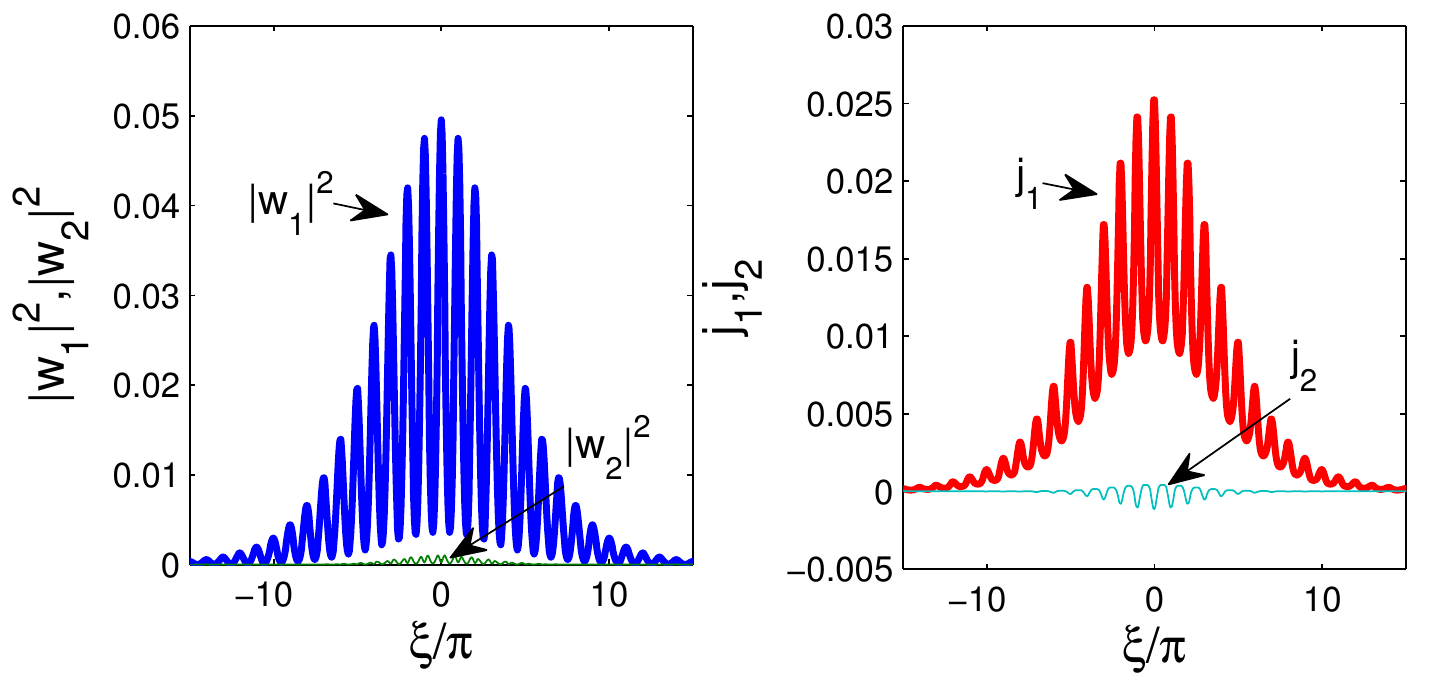}}
\end{center}
\caption{(Color online) An example of a stable \textit{embedded soliton}
with $b=-0.6$, indicated by the black circle in Fig. \protect\ref%
{fig:branchband} inside the SH band. This solution belongs to the branch of
fundamental solitons that bifurcates from $b_{1,-}^{(0)}$ in the second
finite gap. The parameters are $\protect\alpha =0.6$, $V_{1}=V_{2}=1$ and $%
q=0$.}
\label{fig:bandsol}
\end{figure}
We have found that the GS branches extend into the SH\ band for $\alpha >0.41$.
In Fig.~{\ref{fig:bandsol}} we show a typical example of stable ES. In particular is possible to see that both $|w_{1,2}|^2$ decay rapidly despite being in the SH band.

Note that no ESs were found for the conservative version of the present
system, with $\alpha =0$ \cite{Moreira1,Kartashov}. Embedded solitons
were found in Ref. \cite{embed} in the conservative model without the
potential ($V_{1}=V_{2}=0$), but with cubic nonlinear terms added to the
equations, otherwise only \textit{quasi-solitons} can be found, with
non-vanishing tails at $|\xi |\rightarrow \infty $ \cite{Dave}. Furthermore,
in the conservative system the ESs were found only at discrete values of $b$.
\begin{figure}[tbp]
\begin{center}
\scalebox{0.6} {\includegraphics{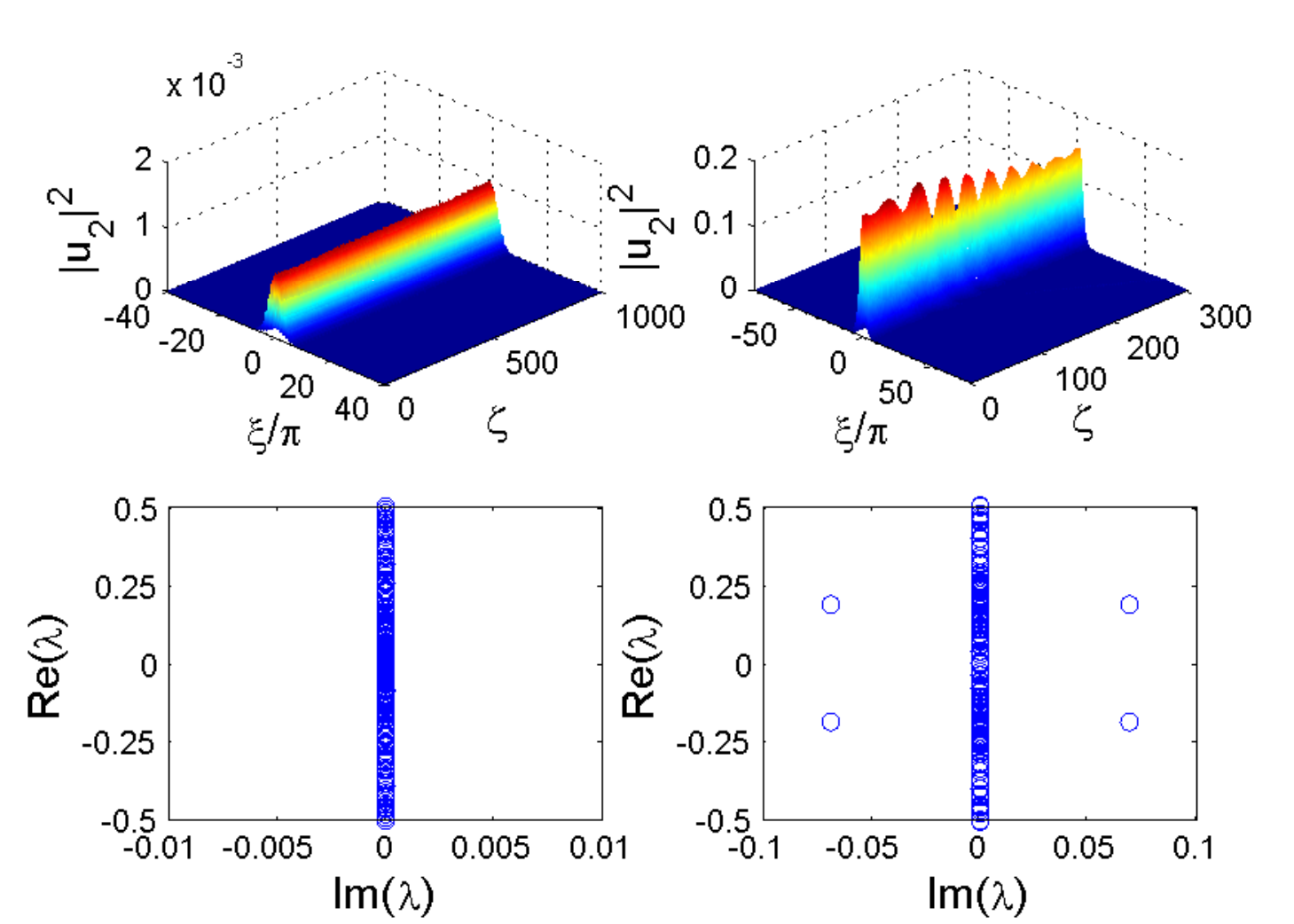}}
\end{center}
\caption{(Color online) Top plots: The evolution of two GS solutions with $%
10\%$ of amplitude random perturbations in Case 3 [see Eq. (\protect\ref%
{case3})] in the second finite gap. Left panel has $b=-0.6$ and is stable.
The right panel corresponds to unstable evolution of a solution with $%
b=-0.65$. The corresponding eigenvalues of small perturbations are shown in
the lower panels. The parameters of the structure are $V_{1}=V_{2}=1$, $%
\protect\alpha =0.6$ and $q=0$.}
\label{fig:propband}
\end{figure}
\begin{figure}[tbp]
\begin{center}
\scalebox{0.5} {\includegraphics{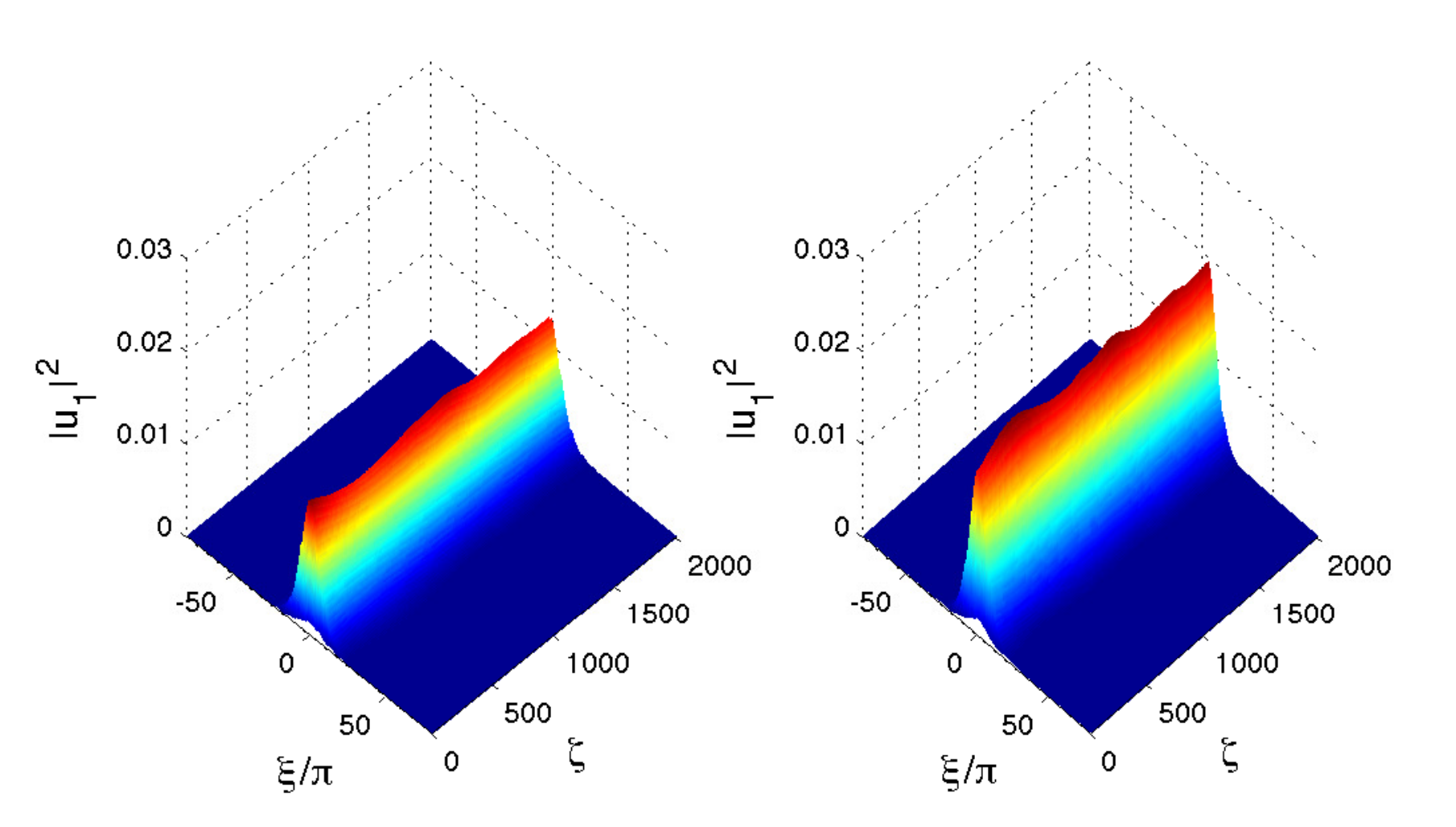}}
\end{center}
\caption{(Color online) An example of the stable evolution of the GS with $%
b=-0.6$, indicated by the black circle in Fig. \protect\ref{fig:branchband},
inside the SH band pertaining to the fundamental branch that bifurcates from
$b_{1,+}^{(1)}$, in the third finite gap. In the left panel the initial
condition is $u_{1,2}(\protect\xi ,0)=0.95\cdot w_{1,2}(\protect\xi )$ and
in the right panel it is $u_{1,2}(\protect\xi ,0)=1.05\cdot w_{1,2}(\protect%
\xi )$. In both cases, the soliton is stable. The parameters are $\protect%
\alpha =0.6$, $V_{1}=V_{2}=1$ and $q=0$.}
\label{fig:pper}
\end{figure}
\begin{figure}[!ht]
\begin{center}
\scalebox{0.64} {\includegraphics{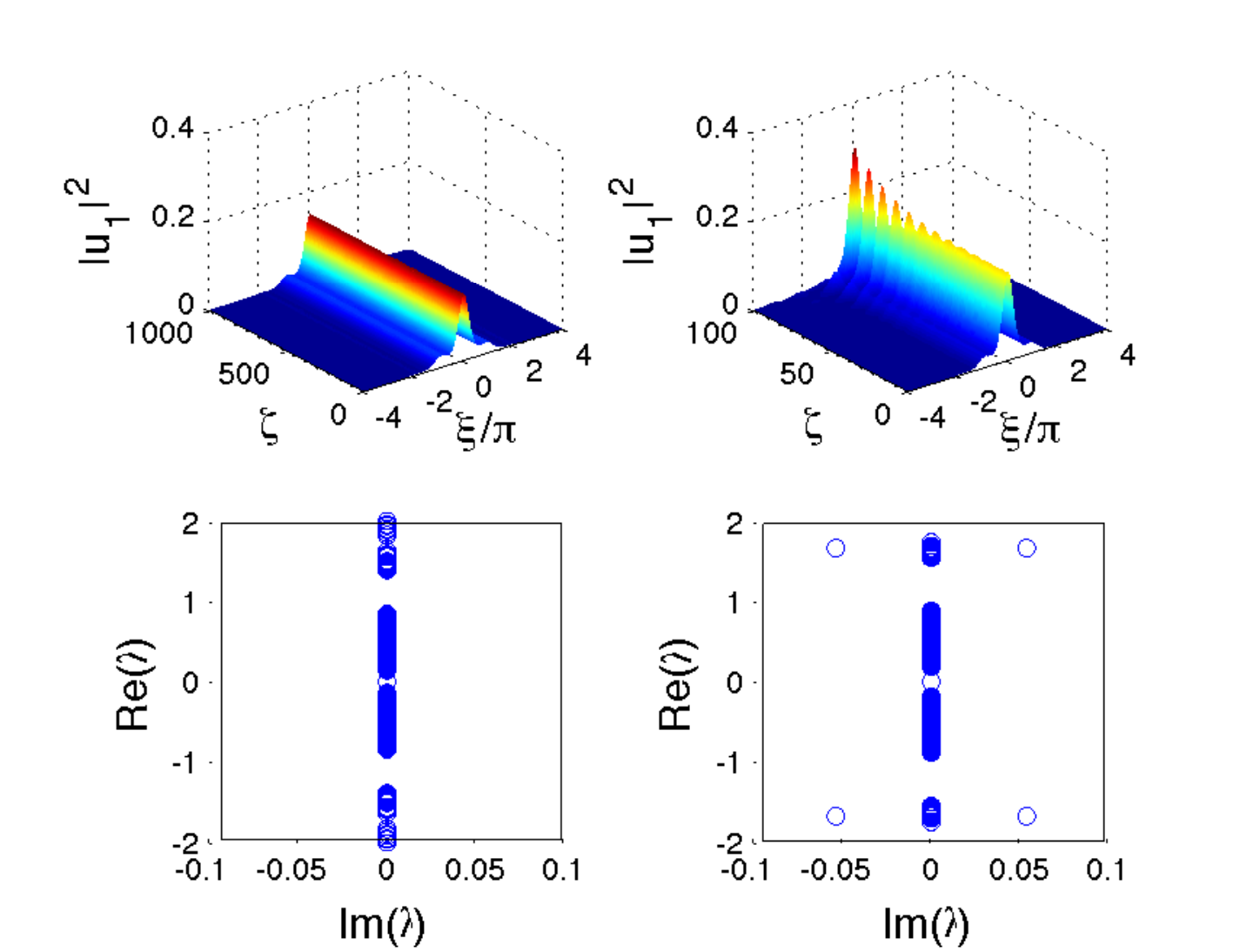}}
\end{center}
\caption{(Color online) Top plots: The evolution of two GS solutions with $%
10\%$ of amplitude random perturbations in Case 3 [see Eq. (\protect\ref%
{case3})] in the semi-infinite gap. Left panel has $b=0.21$ and is stable.
The right panel corresponds to unstable evolution of a solution with $%
b=0.25$. Note that the linearly unstable solution remains localized. The corresponding eigenvalues of small perturbations are shown in
the lower panels. The parameters of the structure are $V_{1}=V_{2}=1$, $%
\protect\alpha =0.7$ and $q=-0.5134$.}
\label{fig:propinfFF}
\end{figure}
A noteworthy feature of ESs in the present system is that a part of their
family is stable, as seen in Fig. \ref{fig:branchband} and in the left panels of in Fig. \ref{fig:propband} while in the
conservative system the isolated ES is \textit{semi-stable} (in Ref. \cite%
{embed}, the ES was stable against perturbations that increased the total
power, but unstable against those which decreased it). The unstable
perturbations in the semi-stable conservative system grow \textit{%
sub-exponentially} [in fact, as $\zeta ^{2}$, rather than as $\exp \left(
\mathrm{const}\cdot \zeta \right) $].
%FIGURA
In our system, the gain component
supplies the power and helps to stabilize perturbed solitons, see Fig. \ref%
{fig:pper}.
Instability, when it appears, is due to the emergence of quartets of complex eigenvalues, as is possible to see in the right panels of Fig. \ref{fig:propband}. The propagation of perturbed unstable solution revealed that the decay is oscillatory.
%FIGURA
We also mention that in Ref. \cite{embed1} continuous families of ESs in a
system with a cubic nonlinearity were found for moving solitons in the plane
of $(v,b)$, where $v$ is the soliton's velocity. However, the ES solutions
still formed discrete sets for any given $v$, including the case of the
quiescent solitons, $v=0$, considered here. To the best of our knowledge,
the present system furnishes the first example a continuous branch of ESs in
a system with a purely quadratic nonlinearity, a part of the branch being
stable.
In the semi-infinite gap, the behavior of solitons in Case 3 is similar to
that in Case 1, outlined above, with one or more alternating stable and
unstable intervals, whose lengths depend on the gain-loss strength, $\alpha $. In Fig. \ref{fig:propinfFF} we show examples of stable and unstable evolutions in the semi-infinite gap. The linear stability analysis shows four complex eigenvalues in the case of the unstable solution. Dynamics shows that instability develops as increasing amplitude oscillations.

In all the finite gaps, we have found two regions, one stable, starting at
the bifurcation point, and the other unstable, as one can see in Fig. \ref%
{fig:branches} for values $\alpha=0.9$ and $\alpha=0.95$ and Fig. \ref{fig:branchband}. In Fig. \ref{fig:prop2gFF} we show examples of stable and unstable solutions in the second-finite gap. The linear stability analysis shows that in the unstable solution the eigenvalues responsible for the instability are purely real. Dynamics shows that the amplitude of the perturbed unstable solution grows without oscillations.
\begin{figure}[!ht]
\begin{center}
\scalebox{0.6} {\includegraphics{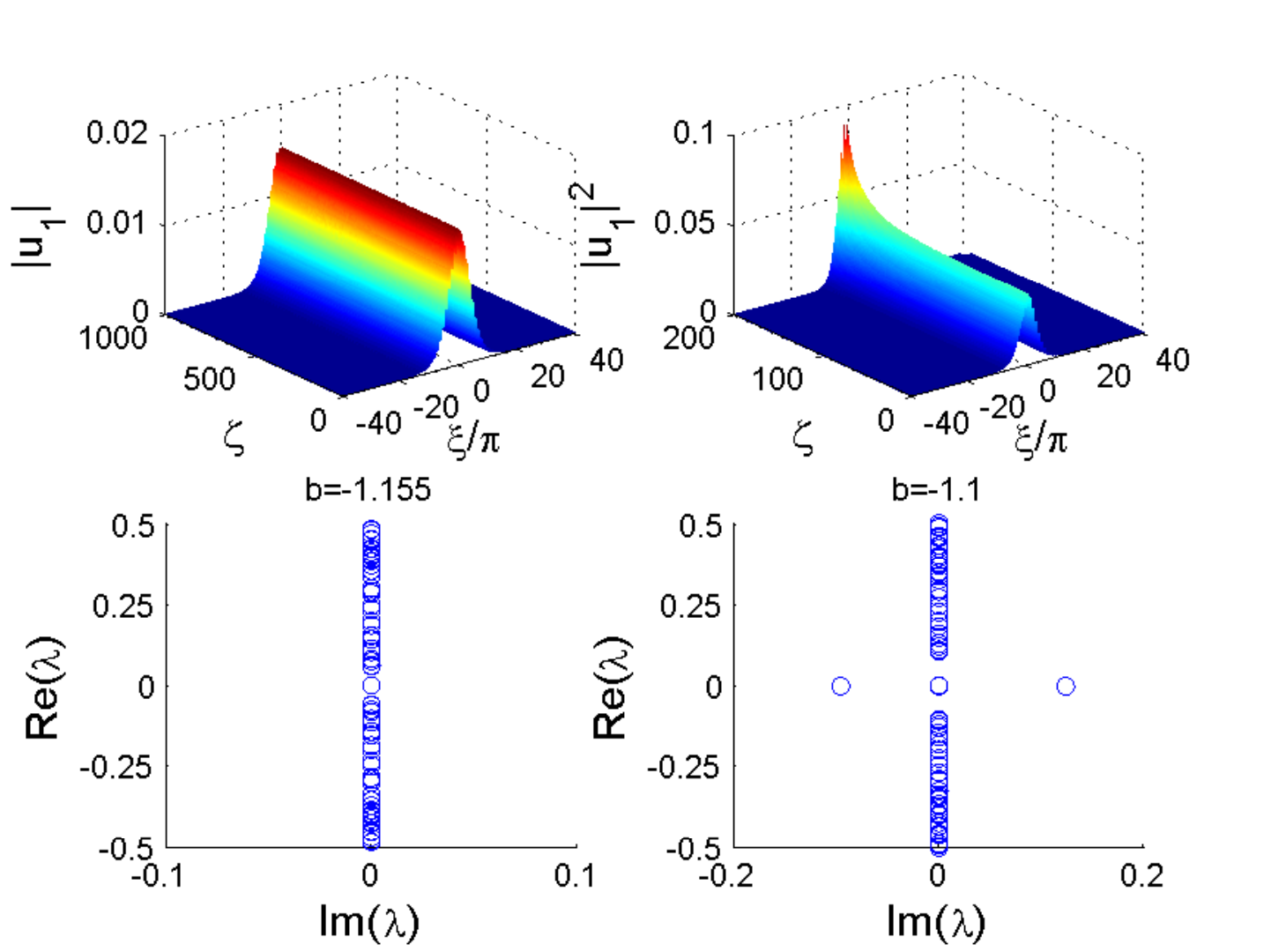}}
\end{center}
\caption{(Color online) Top plots: The evolution of two GS solutions with $%
20\%$ of amplitude random perturbations in Case 3 [see Eq. (\protect\ref%
{case3})] in the second finite gap. Left panel has $b=-1.155$ and is stable.
The right panel corresponds to unstable evolution of a solution with $%
b=-1.1$. Note that the linearly unstable solution remains localized. The corresponding eigenvalues of small perturbations are shown in
the lower panels. The parameters of the structure are $V_{1}=V_{2}=1$, $%
\protect\alpha =0.9$ and $q=0$.}
\label{fig:prop2gFF}
\end{figure}

Lastly, stable solutions have never been found for $|\alpha |>V_{1}$. This
conclusion is qualitatively similar to that made in other nonlinear $%
\mathcal{PT}$ systems, where solitons do not exists above a critical level
of the gain-loss coefficient \cite{Musslimani2008,dual}.

\section{Conclusion}

In this work we have introduced the model combining the linear $\mathcal{PT}$%
-symmetric part and the $\chi ^{(2)}$ nonlinearity. The $\mathcal{PT}$ terms
are represented by the complex potential acting on the FF
(fundamental-frequency) component, whose imaginary part, accounting for the
spatially separated and mutually balanced gain and loss, is, as usual, the
odd function of the coordinate. The potential acting on the SH
(second-harmonic) wave is assumed to be purely real. The complex linear
potential gives rise to the corresponding bandgap spectrum. Solutions for
solitons were looked for in the semi-infinite and finite gaps, starting from
the bifurcation which gives rise to such solitons at edges of the respective
gap. Families of the solitons have been thus constructed, and their
stability was investigated by means of the linearization and direct
simulations alike. While the system contains several parameters, we have
primarily focused on effects produced by the variation of the amplitude of
the imaginary part of the potential, which is specific to the $\mathcal{PT}$%
-symmetric system. A noteworthy result is that the present system may
support of continuous family of solitons embedded into the continuous
spectrum of the SH component, and a part of the family of such embedded
solitons is stable. The analysis has been reported, chiefly, for the most
physically relevant case of equal effective amplitudes of the real
potentials acting on the FF\ and SH waves. In addition, a more exotic case
of the real potential acting solely on the FF component was investigated too
(in Appendix A).

A natural extension of this analysis may be performed for the
two-dimensional version of the $\chi ^{(2)}$ system with the $\mathcal{PT}$%
-symmetric periodic potential. In that case, it may be also interesting to
construct vortex solitons, in addition to the fundamental ones, and
investigate their stability.

\section{Acknowledgments}

V.V.K. acknowledge support of the Funda\c{c}\~ao para a Ci\^encia e a
Tecnologia (Portugal) under the grants PTDC/FIS/112624/2009 and
PEst-OE/FIS/UI0618/2011. F. C. M. was partially supported by Alban fellowship No. E06D100918BR.

\appendix

\section{The system with the ''virtual grating"}

Here we consider the case of $V_{2}=0$, i.e. the periodic potential acting
only on the FF component.
\begin{figure}[!ht]
\begin{center}
\scalebox{0.63} {\includegraphics{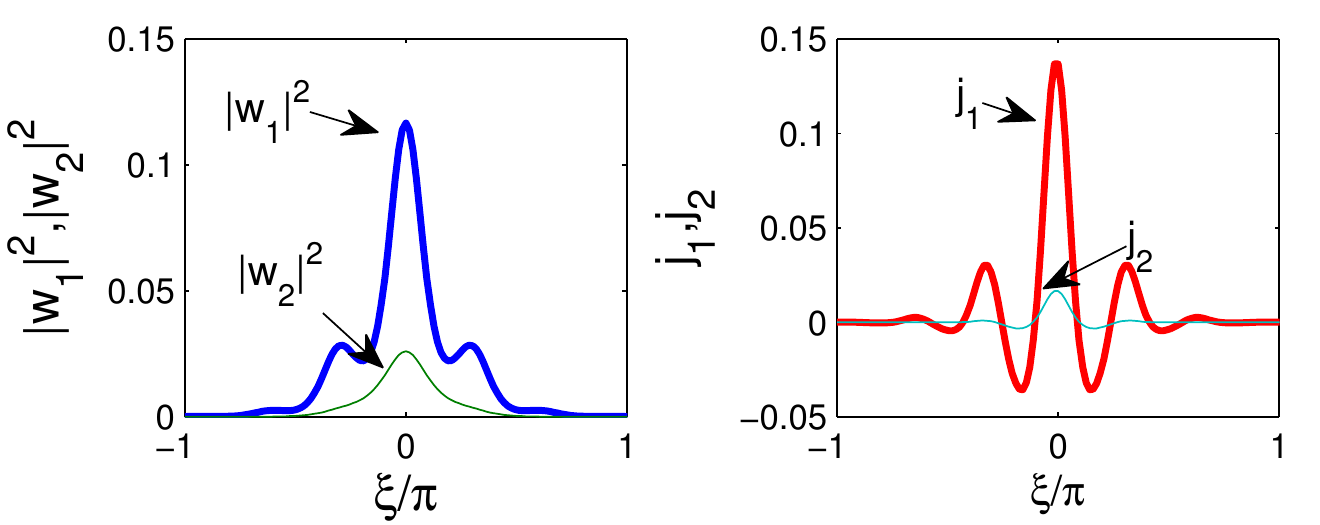}}
\end{center}
\caption{(Color online) The intensities $|w_{1,2}|^2$ and currents $j_{1,2}$ of a stable GS solution
with propagation constant $b=0.2$ pertaining to the semi-infinite gap. The
parameters of the system are $V_{1}=1,V_{2}=0$, $q=0$ and $\protect\alpha %
=0.9$.}
\label{fig:V1-1V20b02alpha09sol}
\end{figure}
\begin{figure}[!ht]
\begin{center}
\scalebox{0.59} {\includegraphics{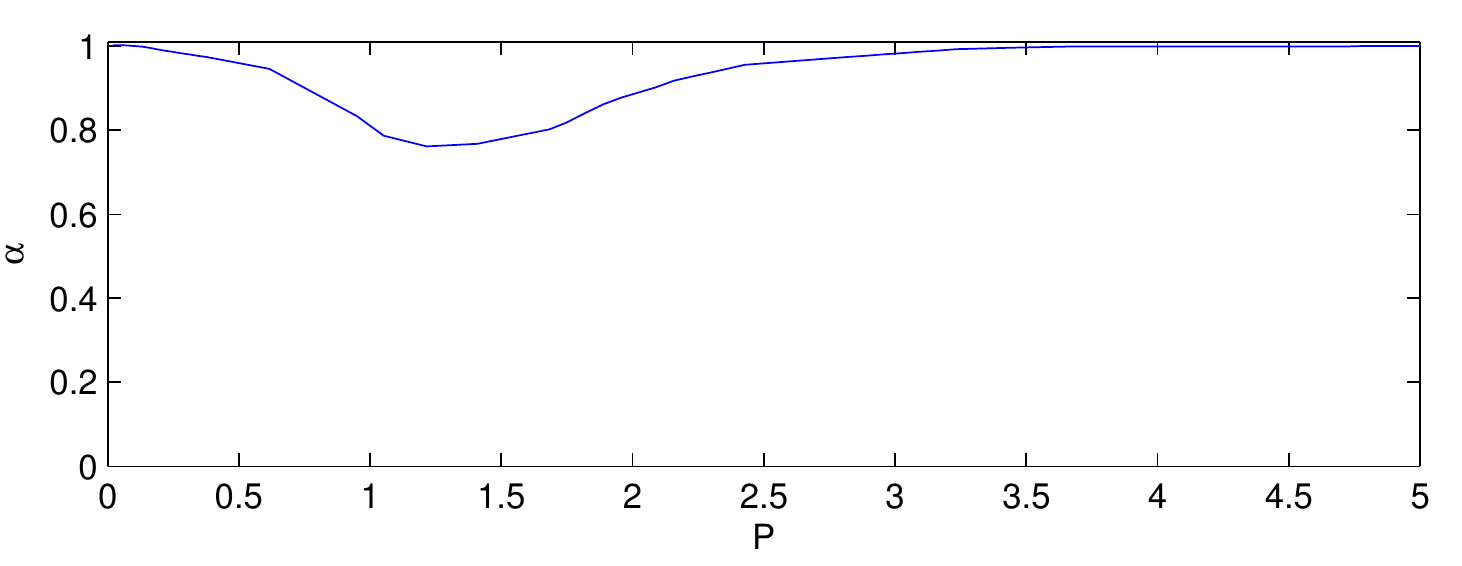}}
\end{center}
\caption{(Color online) The stability boundary in the plane of $\left( P,%
\protect\alpha \right) $, in the system with $V_{1}=1$ and $V_{2}=0$. The
instability area is located above the boundary.}
\label{fig:V1-1V20frontierP}
\end{figure}
A typical example of a stable GS, found as solutions to Eqs. (\ref{stat1})
and (\ref{stat2}) in the absence of the periodic potential acting on the SH,
is displayed in Fig. \ref{fig:V1-1V20b02alpha09sol}. It is seen that its
shape is conspicuously different from that of the solitons found above in
the system with $V_{2}=V_{1}$, cf. Fig. \ref{fig:solutionsP}.

The analysis of the stability of solitons in Eqs. (\ref{final1}) and (\ref%
{final2}) in the case of $V_{2}=0$ reveals a stability boundary, shown in in
Fig. \ref{fig:V1-1V20frontierP}, which is qualitatively similar to its
counterparts presented above for the system with $V_{2}=V_{1}$, cf. Fig. \ref%
{fig:stability1}. It particular, the instability area appears for values of $%
\alpha $ above a certain threshold. However, the difference is that only one
instability interval exists in this case, and the threshold for its
appearance, $\alpha \approx 0.75$, is higher than in the system where the
periodic potential acts on both components.

\section{The case of a purely imaginary potential}

Here we consider the limit case of the system when the potential in Eq. (\ref%
{final1}) is purely imaginary, and no potential appears in Eq. (\ref{final2}%
), i.e., $V_{1}=V_{2}=0$.
\begin{figure}[!ht]
\begin{center}
\scalebox{0.63} {\includegraphics{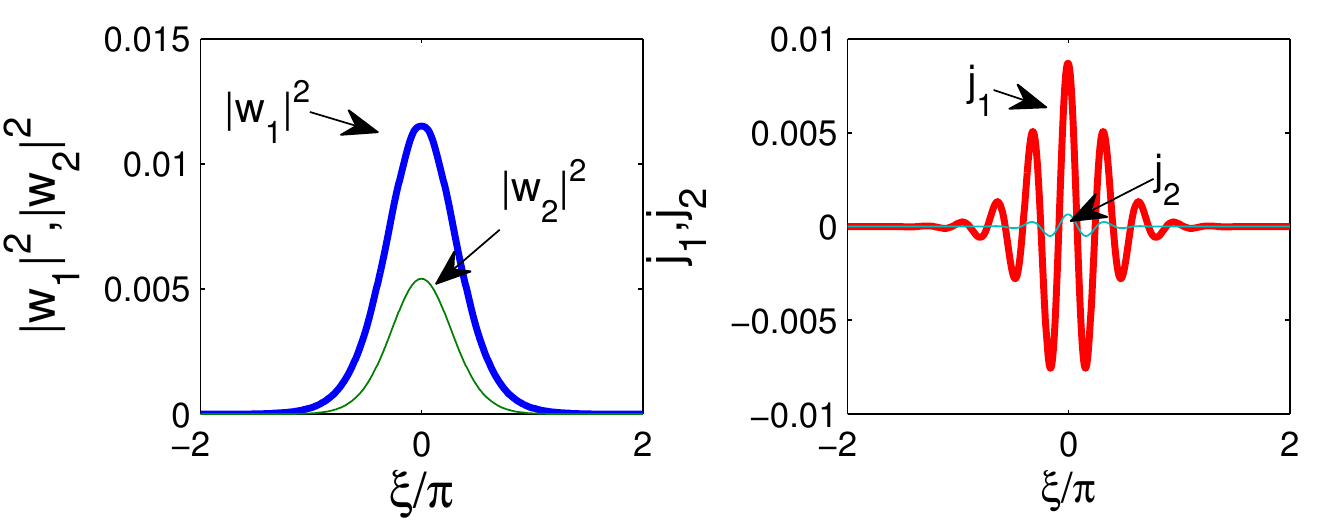}}
\end{center}
\caption{(color online) The intensities $|w_{1,2}|^2$ and currents $j_{1,2}$ of a GS with $b=0.06$, in
the system with $V_{1}=V_{2}=0$ (no real potential) and $q=0$, $\protect%
\alpha =0.5$.}
\label{fig:V10V20b006alpha05sol}
\end{figure}
A typical example of the soliton found in this
case is shown in Fig. \ref{fig:V10V20b006alpha05sol}.

In this case, all the solitons are unstable at $\alpha >0$. The respective
instability growth rate being rather small, Fig. \ref{fig:V10V20stab} shows
that the growth of the instability in direct simulations starts abruptly, as
the instability eigenvalues are purely imaginary, see Eq. (\ref{pert}).
\begin{figure}[!ht]
\begin{center}
\scalebox{0.55} {\includegraphics{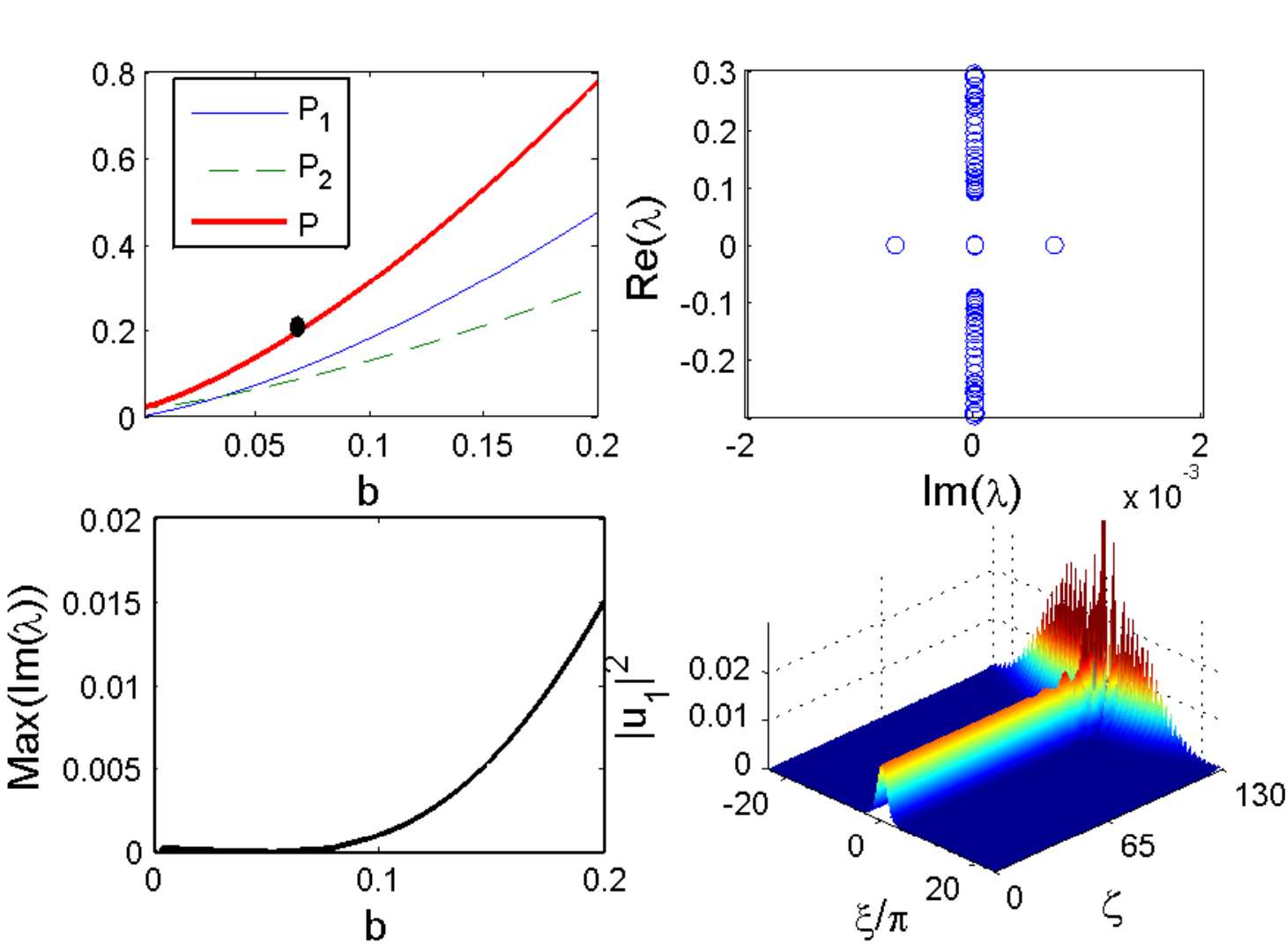}}
\end{center}
\caption{(Color online) Solitons in the system with $V_{1}=V_{2}=q=0$ and $%
\protect\alpha =0.5$. The upper left panel: The
power-vs.-propagation-constant ($b$) branch. The lower left panel: The
instability eigenvalue, $\protect\lambda $, with the largest imaginary part,
as a function of $b$. The upper right panel: Stability eigenvalues for the
soliton with $b=0.06$, the instability being accounted for by a pair of
small purely imaginary eigenvalues. The lower right panel: The unstable
propagation of the soliton randomly perturbed at the 1\% amplitude level.}
\label{fig:V10V20stab}
\end{figure}

\end{subequations}

\end{document}